\documentclass[10pt]{article}

\usepackage{a4wide}

\usepackage{amsmath}
\usepackage{latexsym}

\newcommand{\dd}{{\rm{d}}} 
\newcommand{\R}{\varrho}       
\newcommand{\RR}{\rho}       
\newcommand{\rovno}{\!\!\!\!& = &\!\!\!\!}
\newcommand{\erovno}{\!\!\!\!& = &\!\!\!\!}
\newcommand{\df}{{\phi}} 

\newcommand{\K}{{\cal K}}                   

 \newcommand{\ci}{{\rm i}}                                  
 \newcommand{\Rea}[1]{{\,{\mathrm{Re}\,}#1}\,}              
 \newcommand{\Ima}[1]{{\,{\mathrm{Im}\,}#1}\,}              

 \newcommand{\boldk}{\mbox{\boldmath$k$}}                   
 \newcommand{\boldl}{\mbox{\boldmath$l$}}                   
 \newcommand{\boldm}{\mbox{\boldmath$m$}}                   
 \newcommand{\bboldm}{\bar{\mbox{\boldmath$m$}}}            

\begin{document}

\title{Algebraic classification of Robinson--Trautman spacetimes}

\author{
J.~Podolsk\'y\thanks{{\tt podolsky@mbox.troja.mff.cuni.cz}} \ and \ R.~\v{S}varc\thanks{{\tt robert.svarc@mff.cuni.cz}}\\ \\
Institute of Theoretical Physics, Charles University in Prague,\\
Faculty of Mathematics and Physics, \\
V~Hole\v{s}ovi\v{c}k\'ach 2, 18000 Prague 8, Czech Republic.\\ \\
}

\maketitle

\begin{abstract}
We consider a general class of four-dimensional geometries admitting a null vector field that has no twist and no shear but has an arbitrary expansion. We explicitly present the Petrov classification of such Robinson--Trautman (and Kundt) gravitational fields, based on the algebraic properties of the Weyl tensor. In particular, we determine all algebraically special subcases when the optically privileged null vector field is a multiple principal null direction (PND), as well as all the cases when it remains a single PND. No field equations are a priori applied, so that our classification scheme can be used in any metric theory of gravity in four dimensions. In the classic Einstein theory this reproduces previous results for vacuum spacetimes, possibly with a cosmological constant, pure radiation and electromagnetic field, but can be applied to an arbitrary matter content. As non-trivial explicit examples we investigate specific algebraic properties of the Robinson--Trautman spacetimes with a free scalar field, and also black hole spacetimes in the pure Einstein--Weyl gravity.
\end{abstract}

\vfil\noindent
PACS class: 04.20.Jb, 04.50.--h, 04.30.--w, 04.40.Nr


\vfil
\eject

\section{Introduction}\label{sec_intro}

Robinson--Trautman family of spacetimes was discovered more than half a century ago~\cite{RobTra60,RobTra62}, soon after the advent of new powerful techniques and concepts in general relativity such as geometrical optics of null congruences, null tetrad formalism, and related algebraic classification of the Weyl tensor. It immediately became one of the most fundamental classes of exact solutions of Einstein's field equations, enabling us to construct explicit models in black hole physics, theory of gravitational waves, and cosmology. A great effort has been put into investigation of their various properties.

Geometrically, the Robinson--Trautman class is defined by admitting a non-twisting, shear-free and expanding congruence of null geodesics generated by a vector field ${\mathbf{k}}$ (the non-expanding class defines the closely related Kundt geometries \cite{Kundt:1961,Kundt:1962}).
This group of spacetimes contains many important \emph{vacuum, electrovacuum or pure radiation solutions}, including any value of the \emph{cosmological constat} $\Lambda$. In particular, it involves the well-known spherically symmetric black holes (Schwarzchild, Reissner--Nordstr\"om, Schwarzchild--de Sitter, Vaidya), uniformly accelerating black holes (C-metric), arbitrarily moving Kinnersley's or Bonnor's ``photon rockets'', expanding spherical gravitational waves (including sandwich or impulsive waves) propagating on conformally flat backgrounds with maximal symmetry (Minkowski, de~Sitter, anti-de~Sitter), and even their combinations, e.g., radiative spacetimes with $\Lambda$ settling down to spherical black holes. These are of various algebraically special Petrov--Penrose types (D, N, O, III, II). Details and a substantial list of references can be found in chapter~28 of the monograph \cite{Stephani:2003}, or chapter~19 of \cite{GriffithsPodolsky:2009}.

There has also been a growing interest in Robinson--Trautman spacetimes beyond the standard settings of four-dimensional general relativity and classic matter fields. In \cite{PodOrt06} this family was extended to the \emph{Einstein theory in higher dimensions}~${D>4}$ for the case of an empty space (with any~$\Lambda$) or aligned pure radiation, which revealed substantial differences with respect to the usual ${D=4}$ case. Aligned electromagnetic fields were also incorporated into the Robinson--Trautman higher-dimensional spacetimes within the Einstein--Maxwell theory \cite{OrtaggioPodolskyZofka:2008} (including the Chern--Simons term for odd $D$) and even for more general $p$-form Maxwell fields \cite{OrtaggioPodolskyZofka:2015}.

Absence of \emph{gyratons} (null fluid or particles with an internal spin) in the Robinson--Trautman class of any $D$ was proved in \cite{SvarcPodolsky:2014}. In fact, it was demonstrated that in four dimensions the off-diagonal metric components do not encode the angular momentum of some gyraton but directly determine two independent amplitudes of the Robinson--Trautman exact gravitational waves.

Moreover, new explicit solutions of this type in the Einstein gravity in ${D=4}$ were found and studied, namely Robinson--Trautman solutions with minimally coupled \emph{free scalar field} \cite{TahamtanSvitek:2015} and with electromagnetic field satisfying equations of \emph{nonlinear electrodynamics} \cite{TahamtanSvitek:2016}.

Very recently, a remarkable class of static, spherically symmetric solutions representing black holes in the \emph{Einstein--Weyl gravity} (with higher derivatives) was presented in \cite{LuPerkinsPopeStelle:2015,LuPerkinsPopeStelle:2015b}. It was shown numerically that such a class \emph{contains further black-hole solutions over and above the Schwarzschild solution}. As we will demonstrate below, this also belongs to the Robinson--Trautman class of spacetimes.

Motivated by all these works, we now wish to present a \emph{complete algebraic classification of the four-dimensional} Robinson--Trautman (and Kundt) geometries. As far as we know, this has not been done before because the classic works summarized in \cite{Stephani:2003,GriffithsPodolsky:2009} remained constrained to (electro)vacuum or pure radiation solutions of Einstein's equations which are \emph{algebraically special} due to the celebrated Goldberg--Sachs theorem \cite{GoldbergSachs:1962} and its generalizations, see section~7.6 of \cite{Stephani:2003}.

To this end we will employ explicit components of the curvature tensors (Riemann, Ricci, and Weyl) for the most general class of non-twisting, shear-free geometries in any dimension ${D\ge 4}$ which we calculated in our previous work. This enabled us to determine possible algebraic types and subtypes of such spacetimes in higher dimensions, based to the multiplicity of the Weyl aligned null directions --- following the classification method  summarized in the review \cite{OrtaggioPravdaPravdova:2013}. The particular case of non-expanding Kundt geometries was investigated in \cite{PodolskySvarc:2013a,PodolskySvarc:2013b} while the inclusion of expanding Robinson--Trautman geometries was achieved in \cite{PodolskySvarc:2015}, together with the discussion of vacuum solutions of Einstein's field equations.

In this work we will solely concentrate on the most important ${D=4}$ case which exhibits highly specific properties. Above all, the corresponding transverse Riemannian space is 2-dimensional, i.e., conformally flat which considerably simplifies the possible structure of algebraic subtypes. Also, here we will use a different and more convenient choice of the null tetrad in real coordinates and the corresponding ten independent real Weyl scalars of five distinct boost weights. This we will combine with standard Newmann--Penrose formalism employing a complex null tetrad.

Let us also emphasize that in our analysis we will not initially assume any gravitational field equations, so that the ``purely geometrical'' results can be applied in any metric theory of gravity (not just in Einstein's general relativity), and in the presence of any matter field.

First, in section~\ref{sec:geom} we present the general metric form of a non-twisting, shear-free spacetime, and we introduce the Robinson--Trautman geometries. In section~\ref{sec:genralPsi} we define the null tetrad and the corresponding Weyl scalars, both in real and complex formalisms. Explicit form of these Weyl scalars, crucial for the algebraic classification, are given in section~\ref{sec:Weylscalars}. General method of determining algebraic types of four-dimensional spacetimes and the corresponding principal null directions (PNDs) are recalled in section~\ref{sec:PND}. A detailed discussion of all possible algebraically special subtypes of the Robinson--Trautman geometries is contained in sections~\ref{multiplePND} and~\ref{exceptmultiplePNDi} for the cases when the geometrically privileged null vector field $\mathbf{k}$ is a multiple PND or it remains a single PND, respectively. A remark on the special case of Kundt geometries is given in section~\ref{Kundt}. The final section~\ref{examples} is devoted to applications of our general results to several interesting explicit examples, namely the algebraically special spacetimes of the Robinson--Trautman class in Einstein's theory of gravity --- both of the Ricci type~I (vacuum, aligned Maxwell field) and of a general Ricci type (scalar field) --- and the static, spherically symmetric black holes in the pure Einstein--Weyl gravity. Explicit coordinate components of the Christoffel symbols, Riemann, Ricci, and Weyl tensors for a generic non-twisting, shear-free geometry in four dimensions are presented in appendix~\ref{appendixA}.

\section{The Robinson--Trautman geometries}
\label{sec:geom}

In this paper we will investigate the general family of four-dimensional spacetimes admitting a null vector field ${\mathbf{k}}$ that is \emph{non-twisting} (${\omega=0}$), \emph{shear-free} (${\sigma=0}$) but \emph{expanding} (${\Theta\not=0}$). It was shown already in the original seminal work by Robinson and Trautman \cite{RobTra60,RobTra62}  that the metric of such spacetimes can be written in the form
\begin{equation}\label{obecny netwistujici prostorocas}
\dd s^2 = g_{ij}(r,u,x^k)\, \dd x^i\dd x^j+2g_{ui}(r,u,x^k)\, \dd u\dd x^i-2\dd u\dd r+g_{uu}(r,u,x^k)\, \dd u^2 \,,
\end{equation}
where the coordinates are adapted to the optically privileged null vector field. Namely, $r$ is the affine parameter along a congruence of null geodesics generated by ${\mathbf{k}}$ (so that ${\mathbf{k}=\mathbf{\partial}_r}$), the whole manifold is foliated in such a way that ${\mathbf{k}}$ is everywhere tangent (and normal) to hypersurfaces ${u=\,}$const., and at any fixed $u$ and $r$  the two spatial coordinates ${x^k\equiv(x^2, x^3)}$ span the transverse 2-dimensional Riemannian manifold with the metric $g_{ij}$.\footnote{Throughout this paper the indices $i,j,k,l$ label the spatial directions and range from $2$ to ${3}$.}
Note that the nontrivial components of an inverse metric are $g^{ij}$ (inverse of $g_{ij}$), ${g^{ri}=g^{ij}g_{uj}}$, ${g^{ru}=-1}$ and ${g^{rr}=-g_{uu}+g^{ij}g_{ui}g_{uj}}$ (so that $g_{ui}=g_{ij}g^{rj}$ and $g_{uu}=-g^{rr}+g_{ui}g^{ri}$).

By construction, the metric (\ref{obecny netwistujici prostorocas}) is non-twisting with a non-zero shear $\sigma$ and expansion $\Theta$. The requirement that the congruence generated by ${\mathbf{k}}$ is shear-free implies the condition
\begin{equation}
G_{ij}=0\,, \qquad \hbox{where}\qquad G_{ij}=g_{ij,r}-2\Theta g_{ij}
\,, \label{RTcondition}
\end{equation}
which can be readily integrated to
\begin{equation}
  g_{ij}=\R^2(r,u,x^k)\,h_{ij}(u,x^k)\,, \qquad \hbox{where} \qquad \R_{,r}=\Theta \R \,,
\label{IntShearFreeCond}
\end{equation}
that is ${\,\R=\exp\big(\int\Theta(r,u,x^k)\,\dd r\big)}$. Moreover, since any 2-dimensional spatial metric is conformally flat, without loss of generality we can assume
\begin{equation}
  h_{ij}=\delta_{ij} \,,
\label{hijJEdeltaij}
\end{equation}
if such a choice of gauge is convenient.

\newpage
\section{Null tetrad and corresponding Weyl scalars}
\label{sec:genralPsi}

To evaluate the Weyl scalars determining the algebraic structure of the spacetime it is necessary to set up a normalized reference frame. In our notation this consists of two future oriented null vectors, $\boldk$ and $\boldl$, and two perpendicular real spacelike vectors $\boldm_{(i)}$ (standing for $\boldm_{(2)}$ and $\boldm_{(3)}$) which satisfy the normalization conditions
\begin{equation}
\boldk\cdot\boldl=-1 \,, \qquad \boldm_{(i)}\cdot\boldm_{(j)}=\delta_{ij} \,, \qquad \boldk\cdot\boldk=0=\boldl\cdot\boldl \,, \qquad \boldk\cdot\boldm_{(i)}=0=\boldl\cdot\boldm_{(i)} \,. \label{properties of null frame}
\end{equation}
It is most convenient to identify the vector $\boldk$ with the geometrically privileged null vector field ${\mathbf{k}=\mathbf{\partial}_r}$ which generates the non-twisting, shear-free and affinely parameterized geodesic congruence of the spacetime (\ref{obecny netwistujici prostorocas}). The conditions (\ref{properties of null frame}) are then satisfied by the natural choice of the null frame\footnote{An alternative choice used, e.g., in \cite{PodolskySvarc:2015} is
${\boldk=\mathbf{\partial}_r}$, ${{\boldl}=\frac{1}{2}g_{uu}\,\mathbf{\partial}_r+\mathbf{\partial}_u}$, ${{\boldm}_{(i)}=m_{(i)}^i\mathbf{\partial}_i + m_{(i)}^ig_{ui}\,{\partial}_r}$, from which
the null frame (\ref{null_frame}) is obtained by a null rotation ${\tilde{\boldk}=\boldk}$,
${\tilde{\boldl}=\boldl+\sqrt{2}L^i\boldm_{(i)}+|L|^2\boldk}$, ${\tilde{\boldm}_{(i)}=\boldm_{(i)}+\sqrt{2}L_i\boldk}$,
with ${L_i=-\frac{1}{\sqrt{2}}m_{(i)}^{i}g_{ui}}$ (and dropping the tildes).}
\begin{equation}
\boldk=\mathbf{\partial}_r \,, \qquad \boldl=-{\textstyle\frac{1}{2}}g^{rr}\mathbf{\partial}_r+\mathbf{\partial}_u-g^{ri}\mathbf{\partial}_i \,, \qquad \boldm_{(i)}=m_{(i)}^i\mathbf{\partial}_i\,, \label{null_frame}
\end{equation}
where the coefficients $m_{(i)}^i$ are normalized as ${\,g_{ij}\,m_{(k)}^i m_{(l)}^j=\delta_{kl}\,}$, i.e., ${\,m_{(k)}^i m^{(k)j}=g^{ij}}$.

Our aim is to calculate the components of the Weyl tensor in the frame (\ref{null_frame}) and discuss its algebraic properties. We define \emph{real Weyl scalars} with respect to the frame ${\{\boldk, \boldl, \boldm_{(2)},\boldm_{(3)}\}}$ by\footnote{Due to the symmetries of $C_{abcd}$, all other projections onto the frame vectors can be expressed in terms of (\ref{defPsiCoef}).}
\begin{eqnarray}
  \Psi_{0^{ij}}\rovno C_{abcd}\; k^a\, m_{(i)}^b\, k^c\, m_{(j)}^d \,,  \nonumber \\
  \Psi_{1^{i}} \rovno C_{abcd}\; k^a\, l^b\, k^c\, m_{(i)}^d \,, \nonumber \\
  \Psi_{2S}    \rovno C_{abcd}\; k^a\, l^b\, l^c\, k^d \,, \nonumber \\
  \Psi_{2^{ij}}\rovno C_{abcd}\; k^a\, l^b\, m_{(i)}^c\, m_{(j)}^d \,, \label{defPsiCoef} \\
  \Psi_{3^{i}} \rovno C_{abcd}\; l^a\, k^b\, l^c\, m_{(i)}^d \,, \nonumber\\
  \Psi_{4^{ij}}\rovno C_{abcd}\; l^a\, m_{(i)}^b\, l^c\, m_{(j)}^d \,, \nonumber
\end{eqnarray}
where the indices ${\,i,j=2,\,3}$ again correspond to two transverse spatial directions. The symmetries of the Weyl tensor $C_{abcd}$ imply that ${\Psi_{0^{ij}}=\Psi_{0^{ji}}}$, ${\Psi_{2^{ij}}=-\Psi_{2^{ji}}}$, ${\Psi_{4^{ij}}=\Psi_{4^{ji}}}$, and that these $2 \times 2$ matrices are trace-free. We thus have exactly \emph{two} independent components of each boost weight, namely
$\Psi_{0^{22}}$ and $\Psi_{0^{23}}$,
$\Psi_{1^{2}}$ and $\Psi_{1^{3}}$,
$\Psi_{2S}$ and $\Psi_{2^{23}}$,
$\Psi_{3^{2}}$ and $\Psi_{3^{3}}$,
$\Psi_{4^{22}}$ and $\Psi_{4^{23}}$.

In fact, these scalars defined by (\ref{defPsiCoef}) are simply related to ten real components of the classic five \emph{complex Newman--Penrose scalars}
\begin{eqnarray}
  \Psi_{0} \rovno C_{abcd}\, k^a \,m^b \,k^c \,m^d \,, \nonumber\\
  \Psi_{1} \rovno C_{abcd}\, k^a \,l^b \,k^c \,m^d \,, \nonumber\\
  \Psi_{2} \rovno C_{abcd}\, k^a \,m^b \,\bar{m}^c \,l^d \,, \label{NP} \\
  \Psi_{3} \rovno C_{abcd}\, l^a \,k^b \,l^c \,\bar{m}^d \,, \nonumber\\
  \Psi_{4} \rovno C_{abcd}\, l^a \,\bar{m}^b \,l^c \,\bar{m}^d \,, \nonumber
\end{eqnarray}
in the \emph{complex} null tetrad ${\{\boldk, \boldl, \boldm,\bboldm\}}$. Indeed, with the natural identification
\begin{equation}\label{defm4D}
\boldm\equiv{\textstyle\frac1{\sqrt2}}(\boldm_{(2)}-\ci\,\boldm_{(3)})\,,\qquad \bboldm\equiv{\textstyle\frac1{\sqrt2}}(\boldm_{(2)}+\ci\,\boldm_{(3)})\,,
\end{equation}
we immediately obtain the relations
\begin{eqnarray}\label{Psi2D4}
  \Psi_{0} \rovno  \Psi_{0^{22}}-\ci\,\Psi_{0^{23}} \,, \nonumber\\
  \Psi_{1} \rovno  {\textstyle\frac1{\sqrt2}}(\Psi_{1^{2}}-\ci\,\Psi_{1^{3}}) \,, \nonumber\\
  \Psi_{2} \rovno  -{\textstyle\frac{1}{2}}(\Psi_{2S}+\ci\,\Psi_{2^{23}}) \,, \\
  \Psi_{3} \rovno  {\textstyle\frac1{\sqrt2}}(\Psi_{3^{2}}+\ci\,\Psi_{3^{3}}) \,, \nonumber\\
  \Psi_{4} \rovno  \Psi_{4^{22}}+\ci\,\Psi_{4^{23}} \,, \nonumber
\end{eqnarray}
or inversely
\begin{align}\label{D4rel}
 \Psi_{0^{22}}&= -\Psi_{0^{33}} =  \Rea{\Psi_0} \,,  &\Psi_{0^{23}}&=\Psi_{0^{32}} = -\Ima{\Psi_0} \,, \nonumber \\
 \Psi_{1^{2}} &= \sqrt{2}\,\Rea{\Psi_1} \,,  &\Psi_{1^{3}}&= -\sqrt{2}\,\Ima{\Psi_1} \,, \nonumber \\
 \Psi_{2S}    &= -2\Rea{\Psi_2} \,,  &\Psi_{2^{23}}&= -2\Ima{\Psi_2}\,, \\
 \Psi_{3^{2}} &= \sqrt{2}\,\Rea{\Psi_3} \,,  &\Psi_{3^{3}}&=  \sqrt{2}\,\Ima{\Psi_3} \,, \nonumber \\
 \Psi_{4^{22}}&= -\Psi_{4^{33}} =  \Rea{\Psi_4} \,,  &\Psi_{4^{23}}&=\Psi_{4^{32}} =  \Ima{\Psi_4} \,. \nonumber
\end{align}
Clearly, the real scalars (\ref{defPsiCoef}) constructed from the Weyl curvature tensor are (up to a constant rescaling) just the \emph{real and imaginary parts} of the standard Newman--Penrose complex scalars (\ref{NP}).

\section{Weyl scalars for generic Robinson--Trautman geometries}
\label{sec:Weylscalars}
Now the main point is to explicitly express the key Weyl scalars in the null frame (\ref{null_frame}) using their definition (\ref{defPsiCoef}). The Weyl tensor coordinate components for a \emph{completely general} four-dimensional Robinson--Trautman metric (\ref{obecny netwistujici prostorocas}) are summarized in equations (\ref{Weyl rprq})--(\ref{Weyl upuq}) of appendix~A.
Straightforward but lengthy calculation of the respective projections leads to the following Weyl scalars:
\begin{eqnarray}
\Psi_{0^{ij}} \rovno 0 \,, \label{Psi0ij}\\
\Psi_{1^{i}}  \rovno {\textstyle \frac{1}{2}\,m_{(i)}^i \,N_i} \,, \label{Psi1Tj}\\
\Psi_{2S}     \rovno {\textstyle \frac{1}{3}\, S}\,, \label{Psi2s} \\
\Psi_{2^{ij}} \rovno m_{(i)}^i m_{(j)}^j \,F_{ij} \,, \label{Psi2ij}\\
\Psi_{3^{i}}  \rovno {\textstyle \frac{1}{2}\,m_{(i)}^i \,V_i} \,, \label{Psi3Tj}\\
\Psi_{4^{ij}} \rovno {\textstyle m_{(i)}^i m_{(j)}^j \big(W_{ij}-\frac{1}{2}\,g_{ij} W\big)} \,, \label{Psi4ij}
\end{eqnarray}
where
\begin{eqnarray}
 N_i \rovno -{\textstyle \frac{1}{2}}G_{ui,r}+\Theta_{,i} \,, \label{Np}\\
 S   \rovno {\textstyle \frac{1}{2}\!\,^{S}\!R+\frac{1}{2}G_{uu,r}+\frac{1}{2}g^{ij}G_{ui||j}+2g^{ri}N_i-2\Theta_{,u}} \,, \label{P}\\
 F_{ij} \rovno G_{u[i,j]} \,, \label{Fpq}\\
 V_i \rovno {\textstyle \frac{1}{2}\!\,^{S}\!R\, g_{ui} +\frac{1}{2}g^{kl}e_{kl} G_{ui} -\frac{1}{2}g^{rj}G_{ui||j}+g^{rj}G_{uj||i}+\frac{1}{2}G_{ui,u}-\frac{1}{2}G_{uu,i}} \nonumber \\
&& \hspace{-2.0mm} {\textstyle +\frac{1}{2}g^{jk}(g_{ij,u}-g_{ui||j})G_{uk}-g^{kl}(g_{k[i,u||l]}+g_{u[k,i]||l})+\frac{1}{2}g^{rr}N_i} \,, \label{Vp}\\
\quad W_{ij}\rovno  {\textstyle -\frac{1}{2}g_{uu||ij}-\frac{1}{2}g_{ij,uu}+g_{u(i,u||j)}-\frac{1}{2}e_{ij}G_{uu}+\frac{1}{2}g_{uu,(i}G_{j)u}+\frac{1}{2}g_{uu}G_{u(i||j)} } \nonumber\\
&& \hspace{-2.0mm}  {\textstyle +\frac{1}{4}g^{rk}g_{uk}G_{ui}G_{uj}-\frac{1}{2}\,g_{ui}g_{uj}\big[\,^{S}R-g^{kl}\big(G_{uk||l}+\frac{1}{2}G_{uk}G_{ul}\big)\big] } \nonumber \\
&& \hspace{-2.0mm}  {\textstyle + g^{kl}\big[g_{ui}\big(g_{k[j,u||l]}+g_{u[k,j]||l}\big)+g_{uj}\big(g_{k[i,u||l]}+g_{u[k,i]||l}\big) \big] } \nonumber \\
&& \hspace{-2.0mm}  {\textstyle -g^{kl}e_{kl}G_{u(i}g_{j)u}+g^{kl}G_{uk}e_{l(i}g_{j)u}-\frac{1}{2}g^{rk}G_{uk}G_{u(i}g_{j)u} } \nonumber \\
&& \hspace{-2.0mm}  {\textstyle +g^{kl}E_{ki}E_{lj}-g^{rk}E_{k(i}G_{j)u} -\frac{1}{2}g^{rk}\big(g_{u(i}G_{j)u||k}+G_{uk||(i}g_{j)u} \big) }
\,, \label{Wpq}
\end{eqnarray}
and the contraction $W$ is defined as ${\,W = g^{ij}W_{ij}\,}$.
Here we have introduced convenient functions
\begin{eqnarray}
G_{ui} \erovno g_{ui,r}\,-2\Theta g_{ui} \,, \label{Gui}\\
G_{uu} \erovno g_{uu,r}-2\Theta g_{uu} \,,\label{Guu}
\end{eqnarray}
and auxiliary tensor quantities on the transverse Riemannian 2-space
\begin{eqnarray}
e_{ij} \erovno g_{u{(i||j)}}- {\textstyle \frac{1}{2}}g_{ij,u} \,, \nonumber\\
E_{ij} \erovno g_{u{[i,j]}}+ {\textstyle \frac{1}{2}}g_{ij,u} \,, \nonumber\\
g_{ui||j} \erovno g_{ui,j}-g_{uk}\,^{S}\!\Gamma^{k}_{ij} \,, \nonumber\\
g_{ui,u||j} \erovno g_{ui,uj}-g_{uk,u}\,^{S}\!\Gamma^{k}_{ij} \,, \nonumber\\
g_{uu||ij} \erovno g_{uu,ij}-g_{uu,l}\,^{S}\!\Gamma^{l}_{ij}\,, \label{auxiliaryB}\\
g_{i[k,u||j]} \erovno g_{i[k,j],u}+{\textstyle \frac{1}{2}}(\,^{S}\!\Gamma^{l}_{ik}g_{lj,u}-\,^{S}\!\Gamma^{l}_{ij}g_{lk,u}) \,, \nonumber\\
g_{u[j,k]||i} \erovno g_{u[j,k],i}-\,^{S}\!\Gamma^{l}_{ij}g_{u[l,k]}-\,^{S}\!\Gamma^{l}_{ik}g_{u[j,l]} \,.
\nonumber
\end{eqnarray}
Covariant derivative with respect to $g_{ij}$ is denoted by the symbol ${\,_{||}}$. Of course, ${g_{u{[i,j]}}=g_{u{[i||j]}}}$. The symbol ${\,^{S}\!R}$ is the Ricci scalar for the metric $g_{ij}$ of the transverse Riemannian 2-space.

It can be observed that the key functions (\ref{Np})--(\ref{Wpq}) simplify enormously when the \emph{off-diagonal coefficients} $g_{ui}$ of the Robinson--Trautman metric (\ref{obecny netwistujici prostorocas}) \emph{vanish}, that is
\begin{equation}\label{RTgui=0}
\dd s^2 = g_{ij}(r,u,x^k)\, \dd x^i\dd x^j-2\dd u\dd r+g_{uu}(r,u,x^k)\, \dd u^2 \,.
\end{equation}
Indeed, in such a case
\begin{equation}
 g_{ui}=0 \quad\Rightarrow\quad
 G_{ui}=0\,,\quad
 g^{ri}=0\,,\quad
 g^{rr}=-g_{uu}\,,\quad
\label{g_ui=0}
\end{equation}
so that
\begin{eqnarray}
 N_i \rovno \Theta_{,i} \,, \label{NpSPEC}\\
 S   \rovno {\textstyle \frac{1}{2}\!\,^{S}\!R+\frac{1}{2}G_{uu,r}-2\Theta_{,u}} \,, \label{PSPEC}\\
 F_{ij} \rovno 0 \,, \label{FpqSPEC}\\
 V_i \rovno {\textstyle -\frac{1}{2}G_{uu,i}-g^{jk}g_{j[i,u||k]}-\frac{1}{2}g_{uu}N_i} \,, \label{VpSPEC}\\
 W_{ij}\rovno  {\textstyle -\frac{1}{2}g_{uu||ij}-\frac{1}{2}g_{ij,uu}+\frac{1}{4}g_{ij,u} G_{uu} + \frac{1}{4}g^{kl}g_{ik,u}g_{jl,u} }
\,. \label{WpqSPEC}
\end{eqnarray}
Notice that ${\Psi_{2^{ij}}=0}$ due to (\ref{FpqSPEC}), which indicates that the case (\ref{g_ui=0}) is a specific \emph{algebraically distinct subcase} of the Robinson--Trautman geometry.

Let us also recall that it is always possible to assume
\begin{equation}
 g_{ij}=\R^2(r,u,x^k)\,\delta_{ij}\,,
\label{special_metric}
\end{equation}
see (\ref{IntShearFreeCond}) and (\ref{hijJEdeltaij}),
in which case the normalized spacelike vectors $\boldm_{(i)}$ have simple components ${\,m_{(i)}^k=\varrho^{-1}\delta^k_i\,}$ and expressions (\ref{Psi1Tj})--(\ref{Psi4ij}) reduce to
\begin{eqnarray}
\Psi_{1^{i}}  \rovno {\textstyle \frac{1}{2}}\,\varrho^{-1}\,N_i \,, \qquad
\Psi_{2S}     = {\textstyle \frac{1}{3}}\, S\,, \qquad
\Psi_{2^{ij}} = \varrho^{-2}\,F_{ij} \,, \label{sPsi2ij}\\
\Psi_{3^{i}}  \rovno {\textstyle \frac{1}{2}}\,\varrho^{-1}\,V_i \,, \qquad\,
\Psi_{4^{ij}} = {\textstyle\varrho^{-2}\big(W_{ij}-\frac{1}{2}\,g_{ij} W\big)} \,. \label{sPsi4ij}
\end{eqnarray}

Finally, it can be seen from the definitions (\ref{Gui}) and (\ref{Guu}) that --- once the functions $G_{ui}$ and $G_{uu}$ are determined --- the metric coefficients $g_{ui}$ and $g_{uu}$ can be immediately obtained by integration using the relation ${\R_{,r}=\Theta \R\,}$, see (\ref{IntShearFreeCond}), namely
\begin{equation}
 g_{ui}=\R^2\!\int\! \R^{-2}\, G_{ui}\,\dd r\,,
 \qquad\hbox{and}\qquad
 g_{uu}=\R^2\!\int\! \R^{-2}\, G_{uu}\,\dd r\,.
\label{integr_gui_guu}
\end{equation}

\newpage

\section{Determining the algebraic types and principal null directions (PNDs)}
\label{sec:PND}

Let us emphasize that the results (\ref{Psi0ij})--(\ref{Psi4ij}) with (\ref{Np})--(\ref{Wpq}) are valid for \emph{all  Robinson--Trautman geometries} (and Kundt geometries as well, by setting ${\Theta=0}$), without \emph{any} restriction imposed by specific field equations and/or matter content of the spacetime. This enables us to explicitly determine the algebraic type of an arbitrary spacetime of the form (\ref{obecny netwistujici prostorocas}), (\ref{IntShearFreeCond}), and to find the corresponding principal null directions (together with their multiplicity).

First, we immediately observe from~(\ref{Psi0ij}) that ${\Psi_{0^{ij}} = 0}$. This means that \emph{the optically privileged null vector field} ${\mathbf{k}=\mathbf{\partial}_r}$ \emph{is always a principal null direction} of the Weyl tensor, and the algebraic structure is obviously of type~I with respect to the null frame (\ref{null_frame}).

The next question then arises: What is the explicit condition for the Robinson--Trautman geometry to be of type~II, i.e., \emph{algebraically special}, and what is the corresponding double PND? It is well known (see, e.g., sections 4.2, 4.3, 9.3 in~\cite{Stephani:2003} or the explicit algorithm in \cite{dInvernoRusselClark1971}) that such a condition reads ${I^3=27J^2}$, in terms of scalar polynomial invariants constructed from the complex Newman--Penrose scalars $\Psi_A$ as
\begin{equation}
 I=\Psi_0\Psi_4-4\Psi_1\Psi_3+3\Psi_2^2\,, \qquad
 J={\displaystyle \left|\begin{array}{lll}
 \Psi_0 &\Psi_1 &\Psi_2 \\
 \Psi_1 &\Psi_2 &\Psi_3 \\
 \Psi_2 &\Psi_3 &\Psi_4
 \end{array}\right|. }
  \label{scalarI}
\end{equation}
For \emph{any} Robinson--Trautman geometry the Weyl scalars ${\Psi_{0^{ij}}}$ identically vanish, implying ${\Psi_0=0}$. The invariants (\ref{scalarI}) thus reduce to
\begin{equation}
 I=3\Psi_2^2-4\Psi_1\Psi_3\,, \qquad
 J= \Psi_1 (2\Psi_2\Psi_3-\Psi_1\Psi_4)- \Psi_2^3 \,,
  \label{scalarIb}
\end{equation}
so that the condition ${I^3=27J^2}$  explicitly reads
\begin{equation}
 \Psi_1^2\Big[27(\Psi_1^2\Psi_4^2-4\Psi_1\Psi_2\Psi_3\Psi_4+2\Psi_2^3\Psi_4)
 +64\Psi_1\Psi_3^3-36\Psi_2^2\Psi_3^2\Big]=0 \,.
  \label{scalarIbb}
\end{equation}
The Robinson--Trautman spacetime is algebraically special (admits a double PND) if, and only if, the condition (\ref{scalarIbb}) is satisfied. Clearly, there are two distinct possibilities, namely ${\Psi_1=0}$ and ${\Psi_1\not=0\,}$:
\begin{itemize}
\item
In the case ${\Psi_1=0}$, \emph{the optically privileged vector field} ${\mathbf{k}=\mathbf{\partial}_r}$ \emph{is (at least) a double PND} of the Weyl tensor, and its algebraic structure is of type~II with respect to the null frame (\ref{null_frame}).
\item
In the peculiar case ${\Psi_1\not=0}$, the optically privileged null vector field ${\mathbf{k}=\mathbf{\partial}_r}$ \emph{is not} a double principal null direction (it remains a non-degenerate PND), and there exists \emph{another double PND} in the spacetime provided the expression in the square bracket of (\ref{scalarIbb}) vanishes.
\end{itemize}

In the following sections we will systematically analyze both these cases (including all possible subcases) separately. We will also discuss the conditions for the Robinson--Trautman geometry to be of algebraic type III, N, O and D.

Moreover, we will explicitly determine the corresponding four (possibly multiple) principal null directions. Recall (cf. \cite{Stephani:2003,GriffithsPodolsky:2009}) that any PND $\boldk'$ can be obtained by performing a null rotation of the frame (\ref{null_frame}), (\ref{defm4D}) with a fixed null vector $\boldl$, that is
\begin{equation}
\boldk'=\boldk+K\,\bboldm+\bar{K}\,\boldm+K\bar{K}\,\boldl\,,\qquad
\boldl'=\boldl\,,\qquad
\boldm'=\boldm+K\,\boldl\,,
  \label{nullrotation}
\end{equation}
where the parameter $K$ is a root of the equation
${\>\Psi_4K^4+4\Psi_3K^3+6\Psi_2K^2+4\Psi_1K+\Psi_0=0\,}$.
This always has four complex solutions, each corresponding to one of the four PNDs. Of course, for a degenerate root $K$ we obtain a multiple PND $\boldk'$ given by (\ref{nullrotation}).
Since we employ the frame \eqref{null_frame} in which ${\Psi_0=0}$ for any Robinson--Trautman geometry, this quartic equation reduces to
\begin{equation}
K(\Psi_4K^3+4\Psi_3K^2+6\Psi_2K+4\Psi_1)=0\,,
  \label{quarticRT}
\end{equation}
with an obvious root ${K=0}$ corresponding to the optically privileged PND ${\boldk'=\boldk=\mathbf{k}=\partial_r}$.

\newpage

\section{Multiple PND ${\mathbf{k}=\mathbf{\partial}_r}$ and algebraically special subtypes}
\label{multiplePND}

We will first analyze the most important case ${\Psi_1=0}$, for which the key equation (\ref{quarticRT}) reads
\begin{equation}
K^2(\Psi_4K^2+4\Psi_3K+6\Psi_2)=0\,.
  \label{quarticRTspec}
\end{equation}
Since ${K=0}$ is its \emph{double root}, the optically privileged null vector field ${\mathbf{k}=\partial_r}$ is (at least) a \emph{double PND} of the Weyl tensor, and the corresponding Robinson--Trautman spacetime is of \emph{type~II} (or more special).

In view of (\ref{Psi2D4}), (\ref{Psi1Tj}), such a situation occurs if, and only if,
\begin{equation}
\Psi_{1^{i}}=0
\qquad\Leftrightarrow\qquad
N_i=0 \,,
\label{Psi1=0=Ni}
\end{equation}
for both ${i=2}$ and ${i=3}$ (because the spatial vectors $\boldm_{(i)}$ are independent). Using (\ref{Np}), this condition is equivalent to ${G_{ui,r}=2\Theta_{,i}}$. It can be integrated to
\begin{equation}
G_{ui}\equiv f_i\,,
\qquad\hbox{where}\qquad
f_i(r,u,x)= 2\!\int\!\Theta_{,i}\,\dd r+\varphi_i(u,x)
 \,, \label{Gui=fi}
\end{equation}
in which $\varphi_i$ is any function independent of $r$. Consequently,
\begin{equation}
g_{ui}=\R^2\!\int\! \R^{-2}\, f_i\,\dd r\,. \label{deff}
\end{equation}

Moreover, applying the condition ${N_i=0}$ and (\ref{Gui=fi}), the functions (\ref{P})--(\ref{Wpq}) determining the remaining Weyl scalars (\ref{Psi2s})--(\ref{Psi4ij}) simplify to
\begin{eqnarray}
S \rovno {\textstyle \frac{1}{2}\,^{S}\!R +\frac{1}{2}G_{uu,r}+\frac{1}{2}g^{ij}f_{i||j}-2\Theta_{,u}} \,, \label{PII}\\
F_{ij} \rovno f_{[i,j]} \,, \label{FpqII}\\
 V_i \rovno {\textstyle \frac{1}{2}\!\,^{S}\!R\, g_{ui} +\frac{1}{2}g^{kl}e_{kl} f_i -\frac{1}{2}g^{rj}f_{i||j}+g^{rj}f_{j||i}+\frac{1}{2}f_{i,u}-\frac{1}{2}G_{uu,i}} \nonumber \\
&& \hspace{-2.0mm} {\textstyle +\frac{1}{2}g^{jk}(g_{ij,u}-g_{ui||j})f_k-g^{kl}(g_{k[i,u||l]}+g_{u[k,i]||l})} \,, \label{VpII}\\
\quad W_{ij}\rovno  {\textstyle -\frac{1}{2}g_{uu||ij}-\frac{1}{2}g_{ij,uu}+g_{u(i,u||j)}-\frac{1}{2}e_{ij}G_{uu}
+\frac{1}{2}g_{uu,(i}f_{j)}+\frac{1}{2}g_{uu}f_{(i||j)} } \nonumber\\
&& \hspace{-2.0mm}  {\textstyle +\frac{1}{4}g^{rk}g_{uk}f_{i}f_{j}-\frac{1}{2}\,g_{ui}g_{uj}\big[\,^{S}R-g^{kl}\big(f_{k||l}
+\frac{1}{2}f_{k}f_{l}\big)\big] } \nonumber \\
&& \hspace{-2.0mm}  {\textstyle + g^{kl}\big[g_{ui}\big(g_{k[j,u||l]}+g_{u[k,j]||l}\big)+g_{uj}\big(g_{k[i,u||l]}+g_{u[k,i]||l}\big) \big] } \nonumber \\
&& \hspace{-2.0mm}  {\textstyle -g^{kl}e_{kl}f_{(i}g_{j)u}+g^{kl}f_{k}e_{l(i}g_{j)u}-\frac{1}{2}g^{rk}f_{k}f_{(i}g_{j)u} } \nonumber \\
&& \hspace{-2.0mm}  {\textstyle +g^{kl}E_{ki}E_{lj}-g^{rk}E_{k(i}f_{j)} -\frac{1}{2}g^{rk}\big(g_{u(i}f_{j)||k}+f_{k||(i}g_{j)u} \big) }
\,. \label{WpqII}
\end{eqnarray}

\subsection{Type II subtypes with a double PND $\mathbf{k}$}
\label{subtypesofII}

The Robinson--Trautman spacetimes (\ref{obecny netwistujici prostorocas}), (\ref{IntShearFreeCond}) satisfying the condition (\ref{Gui=fi}) are of type II  with (at least) a double PND ${\mathbf{k}=\mathbf{\partial}_r}$. In addition to ${\Psi_{1^i}=0}$, they may admit the following particular \emph{algebraic subtypes} of the Weyl tensor:

\smallskip
\begin{itemize}
\item \emph{subtype} II(a) $\Leftrightarrow$ ${\Psi_{2S}=0}$ $\Leftrightarrow$ ${S=0}$ $\Leftrightarrow$ the metric function $g_{uu}$ satisfies the relation:
\begin{eqnarray}
&& {\textstyle G_{uu,r}=-\,^{S}\!R-g^{ij}f_{i||j}+4\Theta_{,u}} \,. \label{CondIIa}
\end{eqnarray}
This determines the specific dependence of $G_{uu}(r,u,x)$ on the coordinate $r$, which is the affine parameter along the optically privileged null congruence generated by $\mathbf{k}$, and subsequently also the $r$-dependence of $g_{uu}$ via the second equation of (\ref{integr_gui_guu}).

\item \emph{subtype} II(d) $\Leftrightarrow$ ${\Psi_{2^{ij}}=0}$ $\Leftrightarrow$ ${F_{ij}=0}$:
\begin{equation}
f_{[i,j]}=0 \,, \label{CondIId}
\end{equation}
or equivalently ${f_{i||j}=f_{j||i}}$.
Introducing a 1-form ${\df\equiv f_i\,\dd x^i}$ in the transverse ${2}$-dimensional Riemannian space, this condition says that $\df$ is closed (${\dd \df=0}$). By the Poincar\'{e} lemma, on any contractible domain there exists a potential function ${\cal F}$ such that ${\df=\dd{\cal F}}$, that is ${f_i={\cal F}_{,i}}$. In a general case, such ${\cal F}$ exists only \emph{locally}.
\end{itemize}

In general, there are two additional (distinct) principal null directions $\boldk'$ given by (\ref{nullrotation}). The corresponding two parameters $K$ are solutions of the quadratic equation ${\Psi_4K^2+4\Psi_3K+6\Psi_2=0}$, which follows from (\ref{quarticRTspec}), that is explicitly
\begin{equation}
K=\pm\sqrt{4\Big(\frac{\Psi_3}{\Psi_4}\Big)^2-6\frac{\Psi_2}{\Psi_4}} -2\frac{\Psi_3}{{\Psi_4}}\,. \label{kvadrat1}
\end{equation}
The special case ${\Psi_4=0}$ will be discussed in subsections~\ref{typeIIIi} and~\ref{typeD}.

\subsection{Type III with a triple PND $\mathbf{k}$}
\label{subtypesofIII}

The Robinson--Trautman spacetime is of algebraic type~III with respect to the triple PND ${\mathbf{k}=\partial_r}$ if \emph{both independent conditions} (\ref{CondIIa}) and (\ref{CondIId}) are satisfied \emph{simultaneously}. Indeed, in such a case the boost-weight zero Weyl tensor component $\Psi_2$ vanishes, see (\ref{Psi2D4}), and equation (\ref{quarticRTspec}) reduces to
\begin{equation}
K^3(\Psi_4K+4\Psi_3)=0\,.
  \label{quarticRTspecspec}
\end{equation}
Thus, ${K=0}$ is a \emph{triple root}, so that the optically privileged null vector field ${\mathbf{k}=\partial_r}$ is a triply degenerate principal null direction of the Weyl tensor.

There is just one additional PND $\boldk'$ determined by (\ref{nullrotation}) with the parameter $K$ given by
\begin{equation}
K=-4\frac{\Psi_3}{\Psi_4}\,, \label{KIII}
\end{equation}
which is the fourth root of the complex equation (\ref{quarticRTspecspec}). Again, the special case ${\Psi_4=0}$ is left to section~\ref{typeIIIi}. The Weyl scalars ${\Psi_3, \Psi_4}$ entering the above expressions are explicitly determined by equations (\ref{Psi2D4}), (\ref{Psi3Tj}), (\ref{Psi4ij}), in which the structural functions $V_i$ and $W_{ij}$ take the form (\ref{VpII}) and (\ref{WpqII}), respectively, with the two constraints (\ref{CondIIa}), (\ref{CondIId}).

\subsection{Type N with a quadruple PND $\mathbf{k}$}
\label{typeN}

It immediately follows from (\ref{quarticRTspecspec}) that the geometrically privileged PND ${\mathbf{k}=\partial_r}$ becomes quadruply degenerate if, and only if, ${\Psi_3=0}$ (so that ${K=0}$ becomes a quadruple root). In view of (\ref{Psi2D4}), (\ref{Psi3Tj}), this is equivalent to
\begin{equation}
\Psi_{3^{i}}=0
\qquad\Leftrightarrow\qquad
V_i=0 \,,
\end{equation}
for both ${i=2}$ and ${i=3}$. Using (\ref{VpII}) simplified by (\ref{CondIId}), this condition takes the explicit form
\begin{eqnarray}
G_{uu,i} \rovno \,^{S}\!R\, g_{ui} +g^{rj}f_{i||j}+g^{kl}e_{kl} f_i +f_{i,u} \nonumber \\
&& \hspace{-2.0mm} {\textstyle +g^{jk}(g_{ij,u}-g_{ui||j})f_k-2g^{kl}(g_{k[i,u||l]}+g_{u[k,i]||l})} \,. \label{CondIII}
\end{eqnarray}
This is a specific constraint on the spatial derivatives of the function $G_{uu}$, and thus $g_{uu}$.

In such a case, the only remaining Weyl tensor components form a symmetric and traceless ${2\times 2}$ matrix ${\Psi_{4^{ij}}=m_{(i)}^i m_{(j)}^j (W_{ij}-\frac{1}{2}\,g_{ij} W)}$, see (\ref{Psi4ij}), equivalent to the complex Newman--Penrose scalar ${\Psi_{4}=\Psi_{4^{22}}+\ci\,\Psi_{4^{23}}}$. The structural functions ${W_{ij}}$ for such type~N geometries are explicitly given by (\ref{WpqII}). They directly encode the amplitudes $\Psi_{4^{ij}}$ of the corresponding gravitational waves.

\subsection{Type O geometries}
\label{typeO}

The Weyl tensor vanishes completely if, and only if, \emph{all} the above conditions are satisfied and, \emph{in addition\,}, ${\Psi_4=0}$, equivalent to ${\Psi_{4^{ij}}=0}$. This clearly occurs when
\begin{equation}
W_{ij}={\textstyle\frac{1}{2}}\,g_{ij} W\,, \label{CondO}
\end{equation}
with ${W=g^{ij}W_{ij}}$, which is a specific restriction on the functions $W_{ij}$ of (\ref{WpqII}).

\subsection{Type III$_i$ with a triple PND $\mathbf{k}$ and a PND $\boldl$}
\label{typeIIIi}

Let us now investigate the special case ${\Psi_4=0}$ forbidden in expression (\ref{kvadrat1}), and for which (\ref{quarticRTspec}) reduces just to a cubic equation. It can immediately be seen from the definitions (\ref{defPsiCoef}) that ${\Psi_{0^{ij}}\leftrightarrow\Psi_{4^{ij}}}$ and ${\Psi_{1^{i}}\leftrightarrow\Psi_{3^{i}}}$ under the swap ${\boldk\leftrightarrow\boldl}$ of the null vectors. Consequently, the condition ${\Psi_4=0}$ means that \emph{the null vector $\boldl$ defined in (\ref{null_frame}) is a PND}. Instead of (\ref{nullrotation}) with (\ref{KIII}), that formally diverges in this case, the single separate PND is now given by
\begin{equation}
\boldk'=\boldl=-{\textstyle\frac{1}{2}}g^{rr}\mathbf{\partial}_r+\mathbf{\partial}_u-g^{ri}\mathbf{\partial}_i\,, \label{k'forIIIi}
\end{equation}
in addition to the triply degenerate PND ${\mathbf{k}=\partial_r}$.

\subsection{Type D with a double PND $\mathbf{k}$ and a double PND $\boldl$}
\label{typeD}

In the highly degenerate case when ${\Psi_4=0=\Psi_3}$ and ${\Psi_0=0=\Psi_1}$, \emph{both} the null vectors of the frame (\ref{null_frame}), that is
${\boldk=\mathbf{k}=\mathbf{\partial}_r}$ and ${\boldl=-{\textstyle\frac{1}{2}}g^{rr}\mathbf{\partial}_r+\mathbf{\partial}_u-g^{ri}\mathbf{\partial}_i}$,
are \emph{doubly degenerate principal null directions}.
Such a situation occurs if, and only if,
\begin{equation}
V_i=0 \qquad\hbox{and}\qquad W_{ij}={\textstyle\frac{1}{2}}\,g_{ij} W\,, \label{typeDcond}
\end{equation}
where the functions $V_i$ and $W_{ij}$ are given by (\ref{VpII}) and (\ref{WpqII}).
The only remaining components of the Weyl tensor are thus ${\Psi_{2S}}$  and ${\Psi_{2^{ij}}}$ (of boost-weight zero). If one of them vanishes, we obtain the algebraic subtypes D(a) and D(d), respectively, see the conditions (\ref{CondIIa}) and (\ref{CondIId}).

The explicit conditions (\ref{typeDcond}) look rather complicated to enable a complete integration of the metric functions in the most general case. However, there is a considerable simplification for the Robinson--Trautman geometries with ${g_{ui}=0}$, given by the metric \eqref{RTgui=0}. As can be seen from expressions (\ref{g_ui=0}) and (\ref{PSPEC})--(\ref{WpqSPEC}), all such type~D spacetimes are determined by the conditions
\begin{eqnarray}
 && G_{uu,i} = -2g^{jk}g_{j[i,u||k]}\,, \label{VpSPECD}\\
 && {\textstyle g_{uu||ij}+g_{ij,uu}-\frac{1}{2}g_{ij,u} G_{uu} - \frac{1}{2}g^{mn}g_{im,u}\,g_{jn,u} }\nonumber\\
 &&\hspace{9mm}= {\textstyle\frac{1}{2}g_{ij} g^{kl}\Big(
g_{uu||kl}+g_{kl,uu}-\frac{1}{2}g_{kl,u} G_{uu} - \frac{1}{2}g^{mn}g_{km,u}\,g_{ln,u}
\Big)} \,. \label{WpqSPECD}
\end{eqnarray}
${ F_{ij}=0}$ due to ${f_i=0}$ in this case, see (\ref{FpqSPEC}) and (\ref{FpqII}), therefore such geometries are always of subtype D(d) since (\ref{CondIId}) is automatically satisfied, with the only remaining Weyl component
\begin{eqnarray}
\Psi_2={\textstyle -\frac{1}{2}\Psi_{2S}=-\frac{1}{12}(\,^{S}\!R+G_{uu,r}-4\Theta_{,u})}\,. \label{Psi2D}
\end{eqnarray}
For ${g_{ij}=\R^2(r,u,x^k)\,\delta_{ij}\,}$, the conditions (\ref{VpSPECD}), (\ref{WpqSPECD}) for algebraic type D further simplify to
\begin{eqnarray}
 && \big(G_{uu}-(\log\R^2)_{,u}\big)_{,i} = 0\,,\label{WpqSPECDcfI}\\
 && g_{uu||23}=0=g_{uu||32}\,,\qquad  g_{uu||22}=g_{uu||33}\,. \label{WpqSPECDcfII}
\end{eqnarray}

\subsection{Type~D with a double PND ${\mathbf{k}}$ and a double PND ${\boldk'\not=\boldl}$}
\label{typeDexcpet}

Finally, the special case ${\Psi_1=0}$, ${\Psi_4\not=0}$ of equation (\ref{quarticRTspec}) can take the form
\begin{equation}
K^2\,\Psi_4(K-a)^2=0\,,
  \label{quarticRTfactorb0}
\end{equation}
when the quadratic expression ${\Psi_4K^2+4\Psi_3K+6\Psi_2}$ is ${\Psi_4(K-a)^2}$
with a \emph{double root} ${K=a}$. This happens if, and only if, the discriminant vanishes, i.e.,
\begin{equation}
3\Psi_2\Psi_4=2\Psi_3^2 \,.
  \label{scalarIbpecexcept}
\end{equation}
It represents type~D Robinson--Trautman geometries with a \emph{doubly degenerate} PND ${\mathbf{k}=\mathbf{\partial}_r}$ (corresponding to the root ${K^2=0}$) and \emph{another doubly degenerate} PND $\boldk'$ (corresponding to the root ${(K-a)^2=0}$) given by \eqref{nullrotation} with
\begin{equation}
K=-2\frac{\Psi_3}{\Psi_4}\,. \label{KDexceptional}
\end{equation}

\section{Exceptional type~II cases when ${\mathbf{k}=\mathbf{\partial}_r}$ is a single PND}
\label{exceptmultiplePNDi}

In this section we will analyze the peculiar case of algebraically special Robinson--Trautman geometries for which the optically privileged null vector field  ${\mathbf{k}=\mathbf{\partial}_r}$ \emph{remains a single (non-degenerate) PND} while there is \emph{another null direction} which is doubly or possibly triply degenerate principal null direction of the Weyl tensor.

As shown in section~\ref{sec:PND}, such a situation occurs if, and only if, ${\Psi_1\not=0}$ and
\begin{equation}
27(\Psi_1^2\Psi_4^2-4\Psi_1\Psi_2\Psi_3\Psi_4+2\Psi_2^3\Psi_4)+64\Psi_1\Psi_3^3-36\Psi_2^2\Psi_3^2=0 \,.
  \label{scalarIbpec}
\end{equation}
According to the value of $\Psi_4$, we distinguish two cases:

\subsection{Case ${\Psi_4=0\,}$: the vector $\boldl$ is a PND}
\label{typeIIspec1}
In the case when ${\Psi_4=0}$, the null vector field
${\boldl=-{\textstyle\frac{1}{2}}g^{rr}\mathbf{\partial}_r+\mathbf{\partial}_u-g^{ri}\mathbf{\partial}_i}$ is a principal null direction (in addition to the single PND ${\mathbf{k}=\partial_r}$),
see subsection~\ref{typeIIIi}. The condition (\ref{scalarIbpec}) for the algebraically special spacetime (i.e., type~II admitting a degenerate PND) simplifies substantially to
\begin{equation}
9\Psi_2^2\Psi_3^2=16\Psi_1\Psi_3^3 \,.
  \label{scalarIbpecpsi4=0}
\end{equation}
There are now 3 possible subcases of such geometries:

\subsubsection{Subcase ${\Psi_3\not=0}$ with a single PND  $\boldl$}
In such a case the principal null directions ${\mathbf{k}=\partial_r}$ and $\boldl$ given by \eqref{k'forIIIi} are both single, so that the remaining distinct PND must be a doubly degenerate. Indeed, the key equation \eqref{quarticRT} reduces to
\begin{equation}
2\Psi_3K^2+3\Psi_2K+2\Psi_1=0\,.
  \label{quarticRTspecExc}
\end{equation}
The discriminant ${9\Psi_2^2-16\Psi_1\Psi_3}$ of this quadratic equation vanishes due to \eqref{scalarIbpecpsi4=0}, so that there is a \emph{double root}
\begin{equation}
K=-\frac{3\Psi_2}{4\Psi_3}\,, \label{KIIspec_a}
\end{equation}
uniquely determining the additional double PND $\boldk'$ via (\ref{nullrotation}).

\subsubsection{Subcase ${\Psi_3=0\,,\,\Psi_2\not=0}$ with a double PND  $\boldl$}
Clearly, the vector field $\boldl$ given by \eqref{k'forIIIi} is now a \emph{doubly degenerate PND}, and the key equation \eqref{quarticRTspecExc} reduces to
${3\Psi_2K+2\Psi_1=0}$. The additional single PND $\boldk'$ is thus determined by (\ref{nullrotation}) with
\begin{equation}
K=-\frac{2\Psi_1}{3\Psi_2}\,. \label{KIIspec_b}
\end{equation}

\subsubsection{Subcase ${\Psi_3=0\,,\,\Psi_2=0}$ with a triple PND  $\boldl$ (type III)}
The only nonvanishing Weyl scalar is $\Psi_1$. This means that the optically privileged null vector field  ${\mathbf{k}=\mathbf{\partial}_r}$ is a single (non-degenerate) PND while the the null vector field ${\boldl=-{\textstyle\frac{1}{2}}g^{rr}\mathbf{\partial}_r+\mathbf{\partial}_u-g^{ri}\mathbf{\partial}_i}$ is triply degenerate principal null direction of the Weyl tensor.

\subsection{Case ${\Psi_4\not=0\,}$: the vector $\boldl$ is not a PND}
\label{typeIIspec2}
This seems to be the most peculiar situation. Although the condition \eqref{scalarIbpec} is now very complicated when we explicitly substitute the structural functions (\ref{Np})--(\ref{Wpq}) using (\ref{Psi2D4}) and (\ref{Psi1Tj})--(\ref{Psi4ij}), it is still possible to determine the corresponding multiple PND, distinct from ${\mathbf{k}=\mathbf{\partial}_r}$.

Indeed, the fundamental quartic equation (\ref{quarticRT}) whose three roots ${K\not=0}$ determine the remaining three PNDs must have the following factorized form
\begin{equation}
K\,\Psi_4(K-a)^2(K-b)=0\,.
  \label{quarticRTfactor}
\end{equation}
By comparing the coefficients of different powers of $K$ in (\ref{quarticRT}) and (\ref{quarticRTfactor}) we obtain three conditions
\begin{equation}
2a+b=A\,,\qquad
a^2+2ab=B\,,\qquad
a^2b=C\,,
  \label{conditions}
\end{equation}
where
\begin{equation}
A=-4\frac{\Psi_3}{\Psi_4}\,,\qquad
B=6\frac{\Psi_2}{\Psi_4}\,,\qquad
C=-4\frac{\Psi_1}{\Psi_4}\,.
  \label{ABC}
\end{equation}
The first two conditions imply ${b=A-2a}$ and thus  ${3a^2-2Aa+B=0}$, so that
\begin{equation}
a={\textstyle\frac{1}{3}(A\pm\sqrt{A^2-3B}\,)}\,,\qquad
b={\textstyle\frac{1}{3}(A\mp2\sqrt{A^2-3B}\,)}\,.
  \label{solution}
\end{equation}
Straightforward calculation now shows that the third condition of (\ref{conditions}) is  automatically satisfied provided the relation \eqref{scalarIbpec} is applied, selecting just one of the possible signs (upper or lower) in \eqref{solution}. For example, when ${A>0, B=0}$ the first relation \eqref{solution} reduces to ${a=\frac{1}{3}(A\pm A)}$. This excludes the lower sign because with ${a=0}$ the condition ${a^2b=C\not=0}$ of \eqref{conditions} can not be satisfied.

We also assume ${b\not=0}$ since the case  ${b=0}$ of \eqref{quarticRTfactor}, implying ${C=0 \Leftrightarrow \Psi_1=0}$, represents type~D Robinson--Trautman geometries discussed in subsection~\ref{typeDexcpet}. Notice that for ${\Psi_1=0}$ the condition \eqref{scalarIbpec} reduces to ${3\Psi_2\Psi_4=2\Psi_3^2\Leftrightarrow A^2=4B}$, which is exactly the condition \eqref{scalarIbpecexcept}.

\subsubsection{Subcase ${a\not=b}$ with a double PND  ${\boldk'\not=\boldl}$ (type II)}
In such a case we have a specific unique solution for the principal null directions: there is a \emph{doubly degenerate} PND ${\boldk'\not=\boldl}$ given by ${K=a\not=0}$, and a different single PND given by ${K=b\not=0}$. These are both distinct from  the optically privileged single PND  ${\mathbf{k}=\mathbf{\partial}_r}$ (and also distinct from ${\boldl=-{\textstyle\frac{1}{2}}g^{rr}\mathbf{\partial}_r+\mathbf{\partial}_u-g^{ri}\mathbf{\partial}_i}$).

\subsubsection{Subcase ${a=b}$ with a triple PND  ${\boldk'\not=\boldl}$ (type III)}
In the special case ${a=b \Leftrightarrow A^2=3B\not=0}$, the fundamental quartic equation \eqref{quarticRTfactor} takes the form
\begin{equation}
K\,\Psi_4(K-a)^3=0\,.
  \label{quarticRTfactorIII}
\end{equation}
Clearly, there is the optically privileged single PND  ${\mathbf{k}=\mathbf{\partial}_r}$ and a \emph{triply degenerate} PND ${\boldk'\not=\boldl}$ given by ${K=a=\frac{1}{3}A}$, that is
\begin{equation}
K=-\frac{4\Psi_3}{3\Psi_4}\,. \label{AexceptIII}
\end{equation}
Such type~III geometries occur if, and only if, ${A^2=3B}$ which is equivalent to
\begin{equation}
8\Psi_3^2=9\Psi_2\Psi_4\,, \label{conditionAexceptIII}
\end{equation}
with $\Psi_4$, $\Psi_3$, $\Psi_2$, $\Psi_1$ all non-vanishing.

\section{The Kundt geometries}
\label{Kundt}

We would like to emphasize at this point that all the conditions and expressions for specific algebraic types of the Weyl tensor presented in previous sections~\ref{sec:Weylscalars}--\ref{exceptmultiplePNDi} are \emph{also valid for the Kundt geometries} with \emph{vanishing expansion} of the non-twisting, shear-free null vector field ${\mathbf{k}=\mathbf{\partial}_r}$: it just suffices to set ${\Theta=0}$. In view of \eqref{RTcondition}, \eqref{Gui}, \eqref{Guu} this immediately implies
\begin{eqnarray}
G_{ij} \erovno g_{ij,r}=0\,,\label{GijKundt}\\
G_{ui} \erovno g_{ui,r}\,, \label{GuiKundt}\\
G_{uu} \erovno g_{uu,r}\,,\label{GuuKundt}
\end{eqnarray}
and \eqref{IntShearFreeCond}, \eqref{hijJEdeltaij} simplify to
\begin{equation}
  g_{ij}(u,x^k)=\R^2(u,x^k)\,\delta_{ij}\,. \label{IntShearFreeCondKundt}
\end{equation}

\section{Application of our results on explicit examples}
\label{examples}
We will now illustrate the usefulness of these general results concerning algebraic classification of Robinson--Trautman geometries on several interesting classes of such spacetimes.

\subsection{Algebraically special spacetimes in Einstein's general relativity}
\label{example1}
Algebraically special spacetimes of the Robinson--Trautman class in Einstein's theory of gravity have been extensively studied for decades since their introduction in the original papers~\cite{RobTra60,RobTra62}. These classic results are summarized---and specific references are given---in the monographs \cite{Stephani:2003,GriffithsPodolsky:2009}, namely in chapters~28 and~19, respectively (see also \cite{OrtaggioPodolskyZofka:2008,OrtaggioPodolskyZofka:2015,SvarcPodolsky:2014,PodolskySvarc:2015} for more recent results).

They include \emph{vacuum} spacetimes, possibly with any value of the \emph{cosmological constant}~$\Lambda$, aligned \emph{electromagnetic field}, or \emph{pure radiation field} (null fluid). Indeed, the Goldberg--Sachs theorem and its generalisations guarantee that all such Robinson--Trautman geometries must be algebraically special, with the optically privileged null vector field ${\mathbf{k}=\mathbf{\partial}_r}$ \emph{(at least) doubly degenerate PND}, that is the case ${\Psi_1=0}$ described in section~\ref{multiplePND}. The corresponding metrics can always be written in the form
\begin{equation}\label{RTEinstein}
\dd s^2 = g_{ij}(r,u,x^k)\, \dd x^i\dd x^j-2\dd u\dd r+g_{uu}(r,u,x^k)\, \dd u^2 \,,
\end{equation}
which is exactly the line element \eqref{obecny netwistujici prostorocas} with ${g_{ui}=0}$, i.e., \eqref{RTgui=0}. In such a case the key functions determining the algebraic structure of the spacetimes take simple explicit forms
\begin{eqnarray}
 N_i \rovno \Theta_{,i} = 0 \,, \label{NpSPEC_Ein}\\
 S\rovno {\textstyle \frac{1}{2}\!\,^{S}\!R+\frac{1}{2}G_{uu,r}-2\Theta_{,u}} \,, \label{PSPEC_Ein}\\
 F_{ij} \rovno 0 \,, \label{FpqSPEC_Ein}\\
 V_i \rovno {\textstyle -\frac{1}{2}G_{uu,i}-g^{jk}g_{j[i,u||k]}} \,, \label{VpSPEC_Ein}\\
 W_{ij}\rovno  {\textstyle -\frac{1}{2}g_{uu||ij}-\frac{1}{2}g_{ij,uu}+\frac{1}{4}g_{ij,u} G_{uu} + \frac{1}{4}g^{kl}g_{ik,u}g_{jl,u} }
\,, \label{WpqSPEC_Ein}
\end{eqnarray}
see \eqref{NpSPEC}--\eqref{WpqSPEC}. Let us also recall, see (\ref{IntShearFreeCond}) and (\ref{hijJEdeltaij}), that it is always possible to assume
\begin{equation}\label{special_metric_Ein}
 g_{ij}=\R^2(r,u,x^k)\,\delta_{ij}\,,
\end{equation}
in which case, using \eqref{auxiliaryB} with the Christoffel symbols ${\,^{S}\!\Gamma^{l}_{ik}}$ for the spatial metric \eqref{special_metric_Ein},
\begin{equation}\label{Vitermsimplified}
g^{jk}g_{j[i,u||k]} = -(\log \R)_{,ui}\,,
\end{equation}
and $g_{uu||ij}$ in \eqref{WpqSPEC_Ein} can also easily be evaluated, yielding
\begin{eqnarray}
 g_{uu||22} \rovno g_{uu,22}-g_{uu,2}\,(\log\R)_{,2}+g_{uu,3}\,(\log\R)_{,3} \,, \label{gu22}\\
 g_{uu||33} \rovno g_{uu,33}+g_{uu,2}\,(\log\R)_{,2}-g_{uu,3}\,(\log\R)_{,3} \,, \label{gu33}\\
 g_{uu||23} \rovno g_{uu,23}-g_{uu,2}\,(\log\R)_{,3}-g_{uu,3}\,(\log\R)_{,2} = g_{uu||32} \,. \label{gu23}
\end{eqnarray}
Moreover, the normalized spacelike vectors $\boldm_{(i)}$ have simple components ${\,m_{(i)}^k=\varrho^{-1}\delta^k_i\,}$, so that the null frame \eqref{null_frame} is now
\begin{equation}
\boldk=\mathbf{\partial}_r \,, \qquad \boldl={\textstyle\frac{1}{2}}g_{uu}\,\mathbf{\partial}_r+\mathbf{\partial}_u \,, \qquad \boldm_{(i)}=\varrho^{-1}\mathbf{\partial}_i\,. \label{null_frame_Ein}
\end{equation}
In this frame, the only non-vanishing Weyl scalars (see expressions (\ref{Psi1Tj})--(\ref{Psi4ij})) are
\begin{eqnarray}
\Psi_{2S}     \rovno {\textstyle \frac{1}{3}}\, S\,, \label{Psi2S_Ein}\\
\Psi_{3^{i}}  \rovno {\textstyle \frac{1}{2}}\,\varrho^{-1}\,V_i \,, \label{Psi3i_Ein}\\
\Psi_{4^{ij}} \rovno {\textstyle\varrho^{-2}\big(W_{ij}-{\textstyle \frac{1}{2}}\,\delta_{ij}\delta^{kl} W_{kl}\big)} \,. \label{Psi4ij_Ein}
\end{eqnarray}
Since
${\ W_{ij}=-\frac{1}{2}g_{uu||ij}+\frac{1}{4}\delta_{ij}\,\big[\big(\R^{-1}(\R^2)_{,u}\big)^2
-2 (\R^2)_{,uu}+(\R^2)_{,u} G_{uu} \big]}$, we clearly have
\begin{eqnarray}
 \Psi_{4^{22}} \rovno {\textstyle \frac{1}{2}}\R^{-2} \big(W_{22}-W_{33}\big)
   = {\textstyle \frac{1}{4}}\R^{-2} \big(g_{uu||33}-g_{uu||22}\big)\,, \label{gu22gen}\\
 \Psi_{4^{23}} \rovno \R^{-2} \,W_{23}
   \hspace{16.1mm} = -{\textstyle \frac{1}{2}}\R^{-2} \, g_{uu||23}\,.  \label{gu23gen}
\end{eqnarray}
Recall that ${\Psi_{4^{33}} = -\Psi_{4^{22}}}$ and ${\Psi_{4^{23}} = \Psi_{4^{32}}}$.

\subsubsection{Spacetimes of the Ricci type I}
\label{subsubRicciIEinstein}
Almost all algebraically special Robinson--Trautman spacetimes studied in general relativity so far have had a special form of the energy-momentum tensor $T_{ab}$ such that in the null frame its \emph{highest boost weight vanishes} --- namely, that it satisfies the condition
\begin{equation}
T_{ab}k^ak^b=T_{rr}=0\,. \label{Ricci_type_I}
\end{equation}
Due to Einstein's equations and the fact that ${g_{rr}=0}$, this immediately implies ${R_{rr}=R_{ab}k^ak^b=0}$, i.e., the spacetimes are of aligned Ricci type~I. In view of the explicit form \eqref{Ricci rr} of this Ricci tensor component, this puts a constraint ${\Theta_{,r}=-\Theta^2}$ on the expansion function which can readily be integrated as ${\Theta=\big(r+\psi(u,x^i)\big)^{-1}}$. Since ${\Theta_{,i}=0}$, see \eqref{NpSPEC_Ein}, the integration function $\psi$ must be independent of the spatial coordinates $x^i$. However, any such function $\psi(u)$ can be removed by the gauge transformation ${r\to r-\psi(u)}$ of the metric \eqref{RTEinstein}. Without loss of generality we thus obtain, using \eqref{IntShearFreeCond},
\begin{equation}
\Theta=\frac{1}{r}\quad \Leftrightarrow \quad  \R=\frac{r}{P(u,x^i)}\,, \label{Thtea_type_I}
\end{equation}
and the key Weyl scalars \eqref{Psi2S_Ein}--\eqref{gu23gen} reduce to
\begin{eqnarray}
\Psi_{2S}     \rovno \frac{1}{6} \big(G_{uu,r}+\!\,^{S}\!R\big)\,, \label{Psi2S_EinI}\\
\Psi_{3^{i}}  \rovno -\frac{P}{4r}\big(G_{uu,i}+2(\log \R)_{,ui}\big) \,, \label{Psi3i_EinI}\\
\Psi_{4^{22}} \rovno \frac{P^2}{4r^2} \big(g_{uu||33}-g_{uu||22}\big)\,, \label{gu22ev}\\
\Psi_{4^{23}} \rovno-\frac{P^2}{2r^2} \, g_{uu||23}\,.  \label{gu23ev}
\end{eqnarray}

For an important large class of Robinson--Trautman \emph{(electro)vacuum spacetimes with} $\Lambda$, the metric coefficient $g_{uu}$ takes the explicit form
\begin{equation}
g_{uu}= -\K+2r\,(\log P)_{,u}+\frac{2m}{r}-\kappa \frac{|Q|^2}{2r^2}+\frac{\Lambda}{3}r^2 \label{guu_electrovacuum}
\end{equation}
(see expressions (28.8), (28.37), (28.78) in \cite{Stephani:2003}, or \cite{OrtaggioPodolskyZofka:2008, OrtaggioPodolskyZofka:2015}). Here
\begin{equation}
\K \equiv \Delta\log P=\frac{\!\,^{S}\!R}{2}\, r^2  \label{Kdef}
\end{equation}
is the Gaussian curvature of the transverse 2-space with the metric
${g_{ij}=(r^2/P^2)\,\delta_{ij} }$, and $\Delta$ is the corresponding Laplace operator
(in fact, ${\,^{S}\!R=\mathcal{R}\,r^{-2}}$, where $\mathcal{R}$ is the Ricci scalar calculated with respect to the $r$-independent part of the spatial metric $g_{ij}$, that is ${P^{-2}\delta_{ij}}$).
The parameter $m$ represents the mass while $Q$ typically represents the charge. In view of \eqref{Guu}, the function ${G_{uu}=g_{uu,r}-(2/r)\,g_{uu}}$ is thus
\begin{equation}
G_{uu} = -2(\log P)_{,u}+\frac{2\K}{r}-\frac{6m}{r^2}+2\kappa \frac{|Q|^2}{r^3}\,. \label{Guu_electrovacuum}
\end{equation}
Putting this into expressions \eqref{Psi2S_EinI}, \eqref{Psi3i_EinI}, and relations \eqref{gu22}--\eqref{gu23} where now ${(\log\R)_{,i}=-(\log P)_{,i}}$ into \eqref{gu22ev}, \eqref{gu23ev}, we finally obtain
\begin{eqnarray}
\Psi_{2S}     \rovno \frac{2m}{r^3}-\kappa\frac{|Q|^2}{r^4}\,, \label{Psi2S_final}\\
\Psi_{3^{i}}  \rovno -\frac{P}{2r}\Big(\frac{\K}{r}-\frac{3m}{r^2}+\kappa \frac{|Q|^2}{r^3}\Big)_{,i} \,, \label{Psi3i_final}\\
\Psi_{4^{22}} \rovno \frac{P^2}{4r^2} \Big(\big(g_{uu,33}-g_{uu,22}\big)-2g_{uu,2}\,(\log P)_{,2}+2g_{uu,3}\,(\log P)_{,3} \Big) \,, \label{Psi4_22_final}\\
\Psi_{4^{23}} \rovno -\frac{P^2}{2r^2}\Big(g_{uu,23}+g_{uu,2}\,(\log P)_{,3}+g_{uu,3}\,(\log P)_{,2}\Big)\,,\label{Psi4_23_final}
\end{eqnarray}
where $g_{uu}$ is given by \eqref{guu_electrovacuum}.

In literature it has been a common approach to use a \emph{complex notation} for the two transverse spatial coordinates ${x^k}$, namely
\begin{equation}
\zeta =\textstyle{\frac{1}{\sqrt{2}}}(x^2+\hbox{i}\, x^3)\,,
\qquad\hbox{so that}\qquad
\partial_\zeta = \textstyle{\frac{1}{\sqrt{2}}}(\partial_2-\hbox{i}\, \partial_3)\,.
 \label{complex}
\end{equation}
The metric \eqref{RTEinstein} thus becomes
\begin{equation}\label{RTEinsteincomplex}
\dd s^2 = 2\frac{r^2}{P^2}  \,\dd\zeta\dd\bar\zeta-2\dd u\dd r
+\left[-\K+2r\,(\log P)_{,u}+\frac{2m}{r}-\kappa \frac{|Q|^2}{2r^2}+\frac{\Lambda}{3}r^2\right] \dd u^2 \,,
\end{equation}
with $P^2(u,\zeta,\bar\zeta)$. The only non-vanishing Weyl scalars in the complex null frame \eqref{defm4D}, \eqref{null_frame_Ein}, that is
\begin{equation}
\boldk=\mathbf{\partial}_r \,, \qquad \boldl={\textstyle\frac{1}{2}}g_{uu}\,\mathbf{\partial}_r+\mathbf{\partial}_u \,, \qquad \boldm=\frac{P}{r}\,\mathbf{\partial}_\zeta\,, \label{null_frame_Ein_complex}
\end{equation}
are immediately obtained using \eqref{Psi2D4} and \eqref{Psi2S_final}--\eqref{Psi4_23_final} as:
\begin{eqnarray}
\Psi_{2}  \rovno -\frac{m}{r^3}+\kappa\frac{|Q|^2}{2r^4} \,, \label{Psi2S_NP}\\
\Psi_{3}  \rovno 
 -\frac{P}{2r^2}\,\K_{,\bar\zeta}+\frac{3P}{2r^3}\,m_{,\bar\zeta}-\frac{\kappa P}{2r^4}\,(|Q|^2)_{,\bar\zeta}\,, \label{Psi3_NP}\\
\Psi_{4} \rovno -\frac{1}{2r^2} \Big(P^2g_{uu,\bar\zeta}\Big)_{,\bar\zeta}\nonumber \\
 \rovno \frac{1}{2r^2} \big(P^2 \K_{,\bar\zeta}\big)_{,\bar\zeta}
 -\frac{1}{r} \big(P^2 (\log P)_{,u\bar\zeta}\big)_{,\bar\zeta}
 -\frac{1}{r^3} \big(P^2 m_{,\bar\zeta}\big)_{,\bar\zeta}
 +\frac{\kappa}{4r^4} \big(P^2 (|Q|^2)_{,\bar\zeta}\big)_{,\bar\zeta}
 \,. \label{Psi4_NP}
\end{eqnarray}
These Newman--Penrose complex scalars are in full agreement with expressions (28.10) and (28.38) presented in \cite{Stephani:2003}.

\subsubsection{Spacetimes of a general Ricci type: scalar field}
\label{subsubRicciGEinstein}
Recently, an interesting Robinson--Trautman solution with minimally coupled free scalar field $\phi$ was found and studied in \cite{TahamtanSvitek:2015}. It satisfies the Einstein equations ${R_{ab}-\frac{1}{2} R\, g_{ab}=T_{ab}}$ where ${T_{ab}=\phi_{,a}\phi_{,b}-\frac{1}{2} g_{ab} g^{cd}\phi_{,c}\phi_{,d}}$ (or, equivalently, ${R_{ab}=\phi_{,a}\phi_{,b}}$ ), and ${\Box\phi=0}$. The explicit metric is
\begin{equation}\label{RTscalar field}
\dd s^2 = \frac{r^2 U^2-C^2}{U\,p^2(x,y)}\, (\dd x^2+\dd y^2)-2\dd u\dd r
-\left[\frac{k(x,y)}{U}+r\frac{U_{,u}}{U}\right] \dd u^2 \,,
\end{equation}
with
\begin{eqnarray}
U(u)  \rovno \gamma\,\exp(\omega^2u^2+\eta\, u) \,, \label{U}\\
\Delta\log p \rovno k\,,\quad \Delta k = 4C^2\omega^2\,, \label{Pandk}\\
\phi(r,u) \rovno \frac{1}{\sqrt{2}} \log\left(\frac{rU-C}{rU+C}\right), \label{phi(ru)}
\end{eqnarray}
where ${C, \gamma, \omega, \eta}$ are positive constants. For ${C=0}$ the scalar field vanishes, ${\phi=0}$, and vacuum spacetime is recovered by solving the standard Robinson--Trautman field equation ${\Delta\Delta\log p=0}$ (with ${m=0}$, see \cite{Stephani:2003,GriffithsPodolsky:2009}). Notice also that ${\phi\to0}$ as ${u\to\infty}$.

In fact, this solution can be rewritten using the gauge transformation
\begin{equation}\label{RTscalar fieldgauge}
u = F(\bar u)\,,\quad
r = \frac{\bar r}{F_{,\bar u}}  \,,\quad
\hbox{where}\ F_{,\bar u}=\sqrt{U}\quad \Rightarrow\
\bar u(u)=\frac{1}{\sqrt\gamma}\!\int\!\exp\!\Big(-\frac{\omega^2}{2}u^2-\frac{\eta}{2}u\Big)\dd u\,,
\end{equation}
after which the metric \eqref{RTscalar field} takes an alternative form
\begin{equation}\label{RTscalar fieldAlter}
\dd s^2 = \frac{\bar r^2 -C^2\,U^{-1}}{p^2(x,y)}\, (\dd x^2+\dd y^2)-2\dd \bar u\dd \bar r
-k(x,y)\,\dd \bar u^2 \,.
\end{equation}
This looks simpler, however at the expense of a more complicated form of the function ${U(\bar u)}$ which is obtained by substituting the transcendent function $u(\bar u)$ from \eqref{RTscalar fieldgauge} into \eqref{U}.

Now, it is obvious from \eqref{phi(ru)} that
\begin{equation}
R_{ab}k^ak^b=R_{rr}=T_{rr}=(\phi_{,r})^2=\frac{2C^2U^2}{(r^2U^2-C^2)^2} \not=0\,. \label{Ricci_type_gen}
\end{equation}
so that the highest boost weight of the scalar field energy-momentum tensor $T_{ab}$ is nonvanishing, and consequently the corresponding  Robinson--Trautman spacetime is of \emph{a general Ricci type}.

Comparing \eqref{RTscalar field} with \eqref{RTEinstein}, \eqref{special_metric_Ein} we infer
\begin{equation}
\R^2(r,u,x,y)=\frac{r^2U^2-C^2}{U\,p^2(x,y)}\,,\qquad
g_{uu}(r,u,x,y)=-\frac{k(x,y)}{U}-r\frac{U_{,u}}{U}\,.
\label{special_metric_scalar}
\end{equation}
The corresponding expansion scalar ${\Theta=\R_{,r}/\R=\frac{1}{2}(\R^2)_{,r}/\R^2}$ is thus
\begin{equation}
\Theta=\frac{r\,U^2}{r^2U^2-C^2}\quad \Rightarrow \quad
\Theta_{,i}=0\,, \quad
\Theta_{,u}=-\frac{2C^2r\,UU_{,u}}{(r^2U^2-C^2)^2}\,, \label{Theta_scalar}
\end{equation}
and since ${\Psi_{1^{i}}=\frac{1}{2}\varrho^{-1}N_i=\frac{1}{2}\varrho^{-1}\Theta_{,i}=0}$, the spacetime is \emph{(at least) of Weyl type II}.

Notice also that for ${C=0}$ we obtain ${\Theta=1/r}$ and recover the vacuum case \eqref{Thtea_type_I}, and the same behaviour is obtained for a general $C$ as ${r\to\infty}$. Due to \eqref{Guu},
\begin{equation}
G_{uu} = \frac{r^2U^2+C^2}{r^2U^2-C^2}\frac{U_{,u}}{U\,}+\frac{2krU}{r^2U^2-C^2}\,. \label{Guu_scalar}
\end{equation}
Evaluating $G_{uu,r}$, $G_{uu,i}$, using expressions \eqref{PSPEC_Ein}--\eqref{gu23gen} where ${(\log \R)_{,ui}=0}$, ${(\log \R)_{,i}=-(\log p)_{,i}}$,
${ g_{ij,u}=(\R^2)_{,u}\,\delta_{ij}}$, and the identity
\begin{equation}
\,^{S}\!R = \frac{2kU}{r^2U^2-C^2}\,,
\label{sR-pomoci_k}
\end{equation}
we obtain
\begin{eqnarray}
\Psi_{2S}     \rovno \frac{2}{3}C^2U\frac{r\,U_{,u}-k}{(r^2U^2-C^2)^2}\,, \label{Psi2S_scalar}\\
\Psi_{3^{i}}  \rovno -\frac{r\,U^{3/2}\,p}{2(r^2U^2-C^2)^{3/2}}\,k_{,i} \,, \label{Psi3i_scalar}\\
 \Psi_{4^{22}} \rovno \frac{p^2}{4(r^2U^2-C^2)} \Big(
   (k_{,22}- k_{,33}) +2k_{,2}\,(\log p)_{,2}-2k_{,3}\,(\log p)_{,3}
 \Big)\,, \label{Psi422_scalar}\\
 \Psi_{4^{23}} \rovno \frac{p^2}{2(r^2U^2-C^2)} \Big( k_{,23}+k_{,2}\,(\log p)_{,3}+k_{,3}\,(\log p)_{,2} \Big)\,.
 \label{Psi423_scalar}
\end{eqnarray}
The corresponding Newman--Penrose scalars \eqref{Psi2D4} are
\begin{eqnarray}
\Psi_{2}  \rovno \frac{1}{3}C^2U\frac{k-r\,U_{,u}}{(r^2U^2-C^2)^2}\,, \label{Psi2_scalar}\\
\Psi_{3}  \rovno -\frac{r\,U^{3/2}\,p}{2(r^2U^2-C^2)^{3/2}}\,k_{,\bar\zeta} \,, \label{Psi3_scalar}\\
\Psi_{4} \rovno \frac{1}{2(r^2U^2-C^2)} \big(p^2k_{,\bar\zeta}\big)_{,\bar\zeta}\,. \label{Psi4_scalar}
\end{eqnarray}
They agree with the results presented in \cite{TahamtanSvitek:2015} with identification  ${\zeta=\frac{1}{\sqrt{2}}(x+\hbox{i}\, y)}$ and ${\Psi_{2}\leftrightarrow-\bar\Psi_{2}}$, ${\Psi_{3}\leftrightarrow-\bar\Psi_{1}}$, ${\Psi_{4}\leftrightarrow-\bar\Psi_{0}}$ due to different choice of the null vectors and the sign convention of the Weyl tensor.\footnote{There are typos in Eq. (5.2) of \cite{TahamtanSvitek:2015}, namely missing factors ${P}$ and ${2}$ in $\Psi_0$, and a missing factor ${1/4}$ in $\Psi_1$.}

These Weyl scalars  can be used for explicit discussion of the possible algebraic types of the Robinson--Trautman spacetimes with free scalar field \eqref{RTscalar field}--\eqref{phi(ru)}. Clearly, the optically privileged vector field ${\mathbf{k}=\mathbf{\partial}_r}$ \emph{is a double PND} of the Weyl tensor. Such type~II spacetimes are fully classified in section \ref{multiplePND}. For ${C\not=0}$, however, ${\Psi_2=0\Leftrightarrow k=0=U_{,u}\,}$. From \eqref{Pandk}, \eqref{U} it then follows that ${\omega^2=0\Rightarrow U(u)=\gamma\,\exp(\eta\, u)}$, and ${U_{,u}=0}$ requires ${\gamma\,\eta=0}$ which does not allow any nontrivial form $U(u)$. Therefore, \emph{there are no type III, N or O solutions of the form} \eqref{RTscalar field}--\eqref{phi(ru)}, i.e., such Robinson--Trautman spacetimes with the free scalar field are of \emph{genuine} type II or D.

The spacetimes are of \emph{type D if, and only if,} ${3\Psi_2\Psi_4=2\Psi_3^2}$, see \eqref{scalarIbpecexcept}. Using \eqref{Psi2_scalar}--\eqref{Psi4_scalar} this reads
${C^2(k-r\,U_{,u})(p^2k_{,\bar\zeta})_{,\bar\zeta}=r^2U^2p^2(k_{,\bar\zeta})^2}$. The coefficients of all powers of $r$ must vanish, so that necessarily ${k_{,\bar\zeta}=0}$. Consequently, the spacetimes are of type D ${\Leftrightarrow k=\,}$const., i.e.,  the transverse 2-space has a \emph{constant Gaussian curvature}.  The only nonvanishing Weyl scalar is
\begin{equation}
\Psi_{2} = \frac{1}{3}C^2U\frac{k-r\,U_{,u}}{(r^2U^2-C^2)^2}\,, \label{Psi2_scalark}
\end{equation}
and the two double degenerate PNDs are ${\boldk=\mathbf{\partial}_r}$, ${\boldl=-\frac{1}{2}(k+rU_{,u})/U\,\mathbf{\partial}_r+\mathbf{\partial}_u }$.
Using the gauge \eqref{RTscalar fieldgauge}, such type D metrics can be rewritten in the form \eqref{RTscalar fieldAlter} with constant $k$. It is a warped-product spacetime, somewhat resembling direct-product (Kundt) type D electrovacuum spacetimes of Pleba\'nski and Hacyan \cite{PlebanskiHacyan:1979}, see \cite{GriffithsPodolsky:2009}.

Notice finally that by setting ${C=0}$ we recover (special) \emph{vacuum} Robinson--Trautman spacetime, with the only nonvanishing Weyl scalars \eqref{Psi3_scalar}, \eqref{Psi4_scalar}
\begin{equation}
\Psi_{3} = -\frac{p}{2r^2U^{3/2}}\,k_{,\bar\zeta} \,, \qquad
\Psi_{4} =  \frac{1}{2 r^2U^2} \big(p^2k_{,\bar\zeta}\big)_{,\bar\zeta}\,. \label{Psi34_scalarC}
\end{equation}
With the gauge transformation \eqref{RTscalar fieldgauge}, implying ${r=\bar r/\sqrt{U}}$, the line element \eqref{RTscalar fieldAlter} now reads
\begin{equation}\label{RTscalar fieldAlterXXX}
\dd s^2 = \frac{\bar r^2}{p^2}\, (\dd x^2+\dd y^2)-2\dd \bar u\dd \bar r-k\,\dd \bar u^2 \,.
\end{equation}
This is the metric \eqref{RTEinsteincomplex} for the case ${P(x,y)}$, ${m=0=Q}$, ${\Lambda=0}$  if we identify ${{\bar r}/{p} = {r}/{P}}$, so that ${p=P\sqrt{U}}$ and  ${k=\K}$. Substituting these relations into \eqref{Psi34_scalarC} we obtain
\begin{equation}
\Psi_{3} = -\frac{P}{2\bar r^2}\,\K_{,\bar\zeta} \,, \qquad
\Psi_{4} =  \frac{1}{2 \bar r^2} \big(P^2\K_{,\bar\zeta}\big)_{,\bar\zeta}\,, \label{Psi34_scalarCB}
\end{equation}
which are exactly the relations \eqref{Psi3_NP}, \eqref{Psi4_NP} after dropping the bar over $r$.
Such vacuum spacetimes are clearly of type III, N, or O.

\subsection{Algebraically general spacetimes in Einstein's general relativity}
\label{example2}
To our knowledge, an exact Robinson--Trautman-type solution of Einstein's field equations of \emph{genuine type I is not known}. The authors would be grateful if anybody brings our attention to an explicit example of such an interesting four-dimensional spacetime.

\subsection{Black holes in the Einstein--Weyl gravity}
\label{example3}

As the last example of non-trivial Robinson--Trautman geometries we will now investigate a remarkable class of static, spherically symmetric solutions representing \emph{black holes in the pure Einstein--Weyl gravity}, presented last year in \cite{LuPerkinsPopeStelle:2015,LuPerkinsPopeStelle:2015b}. It was demonstrated by numerical methods that such a class \emph{contains further black-hole solutions over and above the Schwarzschild solution}.

The action of the Einstein--Weyl gravity contains an additional quadratic curvature term, namely ${I=\int (R-\alpha\, C_{abcd}C^{abcd})\sqrt{-g}\,\dd^4x}$, where $\alpha$ is a constant. The corresponding field equations are then ${R_{ab}-\frac{1}{2}R\,g_{ab}=4\alpha\,B_{ab}}$, where ${B_{ab}=(\nabla^c \nabla^d+\frac{1}{2}R^{cd})C_{acbd}}$ is the trace-free Bach tensor. The static, spherically symmetric ansatz of \cite{LuPerkinsPopeStelle:2015} reads
\begin{equation}\label{Einstein-WeylBH}
\dd s^2 = -h(\bar r)\dd t^2+\frac{\dd \bar r^2}{f(\bar r)}+\bar r^2(\dd \theta^2+\sin^2\theta\,\dd \phi^2) \,,
\end{equation}
where the spatial part can be written, using the standard stereographic representation
\begin{equation}\label{transfRTEinstein-WeylBH}
x^2+\hbox{i}\, x^3 = \sqrt{2}\,\zeta = 2\,\tan\frac{\theta}{2}\exp(\hbox{i}\phi)\,,\quad\hbox{as}\quad
\dd \theta^2+\sin^2\theta\,\dd \phi^2 = \frac{\delta_{ij}\,\dd x^i\dd x^j}{(1+\frac{1}{4}\delta_{kl}\,x^k x^l)^{2}}\,.
\end{equation}
This is \emph{equivalent} to a special case of the Robinson--Trautman metric \eqref{RTEinstein}, \eqref{special_metric_Ein} by performing the coordinate transformation
\begin{eqnarray}
\bar r \rovno \RR(r)\,,    \label{tEWtoRT1}\\
t      \rovno u-\int\!\frac{\dd r}{g_{uu}(r)} \,, \label{tEWtoRT2}
\end{eqnarray}
see \cite{PravdaPravdovaPodolskySvarc:2016}. Indeed, the metric \eqref{Einstein-WeylBH}, \eqref{transfRTEinstein-WeylBH} becomes
\begin{equation}\label{RTEinstein-Weyl}
\dd s^2 = \R^2(r,x^k)\,\delta_{ij}\, \dd x^i\dd x^j-2\dd u\dd r+g_{uu}(r)\, \dd u^2 \,,
\end{equation}
where
\begin{equation}
\R(r,x^k) = \frac{\RR(r)}{1+\frac{1}{4}\delta_{kl}\,x^k x^l}\,,\label{compareEinstein-Weyl1}
\end{equation}
with the identification
\begin{eqnarray}
h(\bar r)\rovno -g_{uu}(r) \,,\label{compareEinstein-Weyl2}\\
f(\bar r) \rovno h(\bar r)\,(\RR_{,r})^2\,.\label{compareEinstein-Weyl3}
\end{eqnarray}

For the \emph{simplest choice} ${\RR(r)=r\>\Rightarrow\> \RR_{,r} = 1}$ we obtain ${\bar r = r}$ and
\begin{equation}\label{RTEinstein-Weyl1}
f=h=-g_{uu}(r)\,.
\end{equation}
The corresponding expansion scalar is ${\Theta=\R_{,r}/ \R =1/r}$, and the Ricci tensor component \eqref{Ricci rr} is thus trivial, ${R_{rr}=R_{ab}k^ak^b=0}$, which means that the spacetime is (at least) of aligned Ricci type~I, cf. expression \eqref{Thtea_type_I}. It is an analogue of the classic \emph{Schwarzschild black hole solution} from the Einstein gravity (${\alpha=0}$), as described in subsection~\ref{subsubRicciIEinstein}. It is well known that such spherically symmetric vacuum spacetime is of Weyl type~D (see expressions \eqref{Psi2S_NP}--\eqref{Psi4_NP} which, for a constant Gaussian curvature $\K$, simplify to ${\Psi_{2}=-m/r^3}$).

Interestingly, as has been demonstarted numerically in \cite{LuPerkinsPopeStelle:2015,LuPerkinsPopeStelle:2015b}, in the pure Einstein--Weyl gravity with quadratic curvature terms (${\alpha\not=0}$), there exists an \emph{additional branch of static, spherically symmetric solutions distinct from the Schwarzschild black holes}. These non-Schwarzschild black holes have
\begin{equation}\label{RTEinstein-Weyl2}
f\not=h \quad \Leftrightarrow\quad \RR_{,r} \not= 1\,,
\end{equation}
i.e, $\R$ given by \eqref{compareEinstein-Weyl1} can not be simply linear in the coordinate $r$ (the affine parameter along the geometrically privileged null congruence of the Robinson--Trautman geometry).
To apply the general results presented in this paper, we can now determine the algebraic type of such solutions.

Clearly, the expansion scalar ${\Theta=\R_{,r}/ \R =\RR_{,r}/ \RR \not =1/r}$. The Ricci tensor component \eqref{Ricci rr} is non-trivial, ${R_{rr}=R_{ab}k^ak^b\not=0}$, which means that the spacetime is of a \emph{general Ricci type}.

The Weyl type follows from explicit expressions \eqref{PSPEC_Ein}--\eqref{gu23gen} which simplify considerably to
\begin{equation}
\Psi_{2S} = {\textstyle \frac{1}{6}}\, (\!\,^{S}\!R+G_{uu,r} )
= \frac{1}{6} \left(\frac{2}{\RR^2}+\Big[\RR^2\Big(\frac{g_{uu}}{\RR^2}\Big)_{,r}\Big]_{,r} \right) , \label{Psi2S_EinWeyl}
\end{equation}
where we have used the fact that the Ricci scalar of the transverse 2-space of a positive constant  curvature is ${\,^{S}\!R=2\K/\RR^2}$ with ${\K=1}$, cf. \eqref{Kdef},
and ${G_{uu}=g_{uu,r}-2(\RR_{,r}/\RR)\,g_{uu}=\RR^2(g_{uu}/\RR^2)_{,r}}$.
The spacetime is clearly of \emph{Weyl type~D}.

The corresponding Newman--Penrose scalar ${\Psi_{2}=-\frac{1}{2}\Psi_{2S}}$ can be rewritten using the relations \eqref{tEWtoRT1}, \eqref{compareEinstein-Weyl2}, \eqref{compareEinstein-Weyl3}, implying ${\partial_r=\sqrt{f/h}\,\partial_{\bar r}}$,  as
\begin{eqnarray}
\Psi_{2} \rovno  \frac{1}{12}\,
\bigg(\!\!-\frac{2}{{\bar r}^2}+\sqrt{\frac{f}{h}}\,\bigg[{\bar r}^2\,\sqrt{\frac{f}{h}}\left(\frac{h}{{\bar r}^2}\right)'\bigg]' \bigg)  \nonumber\\
\rovno \frac{1}{12}
\left(\frac{2}{{\bar r}^2}\Big[-1+\frac{f}{h}\big(h-{\bar r}h'+{\textstyle\frac{1}{2}}{\bar r}^2h''\big)\Big]
-\frac{1}{\bar r}\,\Big(\frac{f}{h}\Big)'\big(h-{\textstyle\frac{1}{2}}{\bar r}h'\big)\right) , \label{Psi2S_EinWeyl_NP}
\end{eqnarray}
where the prime denotes the derivative with respect to ${\bar r}$. For the simpler Schwarzschild-like case \eqref{RTEinstein-Weyl1}, that is ${f=h}$ and ${\bar r = r}$, this reduces to
\begin{equation}
\Psi_{2} = \frac{1}{6\,{\bar r}^2}\big(\!-1+h-{\bar r}h'+{\textstyle\frac{1}{2}}{\bar r}^2h''\big)\,. \label{Psi2S_EinWeyl_NPSchw}
\end{equation}
For ${f=h=1-2m/{\bar r}}$ we obtain ${\Psi_2=-m/r^3}$, in full agreement with expression \eqref{Psi2S_NP}.

Moreover, we observe from \eqref{Psi2S_EinWeyl_NP} that the general black hole spacetime in the Einstein--Weyl gravity is asymptotically flat (${\Psi_2\to 0}$) when ${f\to1}$ and ${h\to\> }$const. as ${\bar r \to \infty}$.

\section{Summary}

We found and described the possible algebraic structures of a general class of non-twisting and shear-free spacetimes in four dimensions \eqref{obecny netwistujici prostorocas}, that is, the complete Robinson--Trautman (and Kundt) family. Our discussion was based on the explicit Weyl scalars \eqref{Psi0ij}--\eqref{Psi4ij} with \eqref{Np}--\eqref{Wpq} which we obtained by projecting the Weyl tensor components onto the most suitable null tetrad.

Generically, such geometries are of Weyl type~I, and the optically privileged null vector field ${\mathbf{k}=\mathbf{\partial}_r}$ is always a principal null direction of the Weyl tensor.

We derived the necessary and sufficient conditions for all possible algebraically special types such that the null direction $\mathbf{k}$  is a multiple PND. These identify the spacetimes of type II, subtypes II(a) and II(d), type III, N, O, III$_i$ and D, see the explicit conditions given in the corresponding subsections of section~\ref{multiplePND}. In the subsequent section~\ref{exceptmultiplePNDi} we also analyzed the exceptional case when the optically privileged null direction ${\mathbf{k}}$ remains a single PND. Such geometries can be of type I, II or III.
For all these algebraic types we found the corresponding four principal null directions.

These conditions can also immediately be applied to non-expanding Kundt geometries, see section~\ref{Kundt}. Moreover, all our results can be used in any metric theory of gravity that admits non-twisting and shear-free geometries.

The field equations impose specific constraints on admissible algebraic types. Therefore, we investigated several examples in section~\ref{examples}. We analyzed (Weyl) algebraically special spacetimes of the Robinson--Trautman class in Einstein's general relativity, namely the Ricci type~I solutions (vacuum spacetimes, possibly with $\Lambda$, aligned electromagnetic field, or pure radiation in subsection~\ref{subsubRicciIEinstein}), and spacetimes of a general Ricci type (free scalar field in subsection~\ref{subsubRicciGEinstein}). Recently identified static, spherically symmetric black holes in the pure Einstein--Weyl gravity were studied in section~\ref{example3}.

\section*{Acknowledgements}
This work has been supported by the grant GA\v{C}R P203/12/0118. R.\v{S}. also acknowledges the support by UNCE~204020/2012 and the Mobility grant of the Charles University.

\appendix
\section{Riemann, Ricci and Weyl tensors}
\label{appendixA}

The Christoffel symbols for the general Robinson--Trautman metric (\ref{obecny netwistujici prostorocas}) are
\begin{eqnarray}
&& {\textstyle \Gamma^r_{rr} = 0} \,, \label{ChristoffelBegin} \\
&& {\textstyle \Gamma^r_{ru} = -\frac{1}{2}g_{uu,r}+\frac{1}{2}g^{ri}g_{ui,r}} \,, \\
&& {\textstyle \Gamma^r_{ri} = -\frac{1}{2}g_{ui,r}+\Theta g_{ui}} \,, \\
&& {\textstyle \Gamma^r_{uu} = \frac{1}{2}\big[-g^{rr}g_{uu,r}-g_{uu,u}+g^{ri}(2g_{ui,u}-g_{uu,i})\big]} \,, \\
&& {\textstyle \Gamma^r_{ui} = \frac{1}{2}\big[-g^{rr}g_{ui,r}-g_{uu,i}+g^{rj}(2g_{u[j,i]}+g_{ji,u})\big]} \,, \\
&& {\textstyle \Gamma^r_{ij} = -\Theta g^{rr}g_{ij}-g_{u(i||j)}+\frac{1}{2}g_{ij,u}} \,, \\
&& {\textstyle \Gamma^u_{rr}=\Gamma^u_{ru}=\Gamma^u_{ri} = 0} \,, \\
&& {\textstyle \Gamma^u_{uu} = \frac{1}{2}g_{uu,r}} \,, \\
&& {\textstyle \Gamma^u_{ui} = \frac{1}{2}g_{ui,r}} \,, \\
&& {\textstyle \Gamma^u_{ij} = \Theta g_{ij}} \,, \\
&& {\textstyle \Gamma^k_{rr} = 0} \,, \\
&& {\textstyle \Gamma^k_{ru} = \frac{1}{2}g^{kl}g_{ul,r}} \,, \\
&& {\textstyle \Gamma^k_{ri} = \Theta\delta^k_i} \,, \\
&& {\textstyle \Gamma^k_{uu} = \frac{1}{2}\big[-g^{rk}g_{uu,r}+g^{kl}(2g_{ul,u}-g_{uu,l})\big]} \,, \\
&& {\textstyle \Gamma^k_{ui} = \frac{1}{2}\big[-g^{rk}g_{ui,r}+g^{kl}(2g_{u[l,i]}+g_{li,u})\big]} \,, \\
&& {\textstyle \Gamma^k_{ij} = -\Theta g^{rk}g_{ij}+\,^{S}\!\Gamma^k_{ij}} \,, \label{ChristoffelEnd}
\end{eqnarray}
where ${\,^{S}\!\Gamma^k_{ij}\equiv\frac{1}{2}g^{kl}(2g_{l(i,j)}-g_{ij,l})}$ are the Christoffel symbols with respect to the spatial coordinates (determining the covariant derivative on the transverse 2-dimensional Riemannian space).

The Riemann curvature tensor components are
\begin{eqnarray}
&& R_{rirj} = {\textstyle -\big(\Theta_{,r}+\Theta^2\big)g_{ij}} \,, \\
&& R_{riru} = {\textstyle -\frac{1}{2}g_{ui,rr}+\frac{1}{2}\Theta g_{ui,r}} \,, \\
&& R_{rikj} = {\textstyle 2g_{i[k}\Theta_{,j]}-2\Theta^2g_{i[k}g_{j]u}+\Theta g_{i[k}g_{j]u,r}} \,, \\
&& R_{ruru} = {\textstyle -\frac{1}{2}g_{uu,rr}+\frac{1}{4}g^{ij}g_{ui,r}g_{uj,r}} \,, \\
&& R_{riuj} = {\textstyle \frac{1}{2}g_{ui,r||j}+\frac{1}{4}g_{ui,r}g_{uj,r}-g_{ij}\Theta_{,u}} \nonumber \\
&& \hspace{15.0mm} {\textstyle -\frac{1}{2}\Theta\big(g_{ij,u}+g_{ij}g_{uu,r}+g_{uj}g_{ui,r}-g_{ij}g^{rl}g_{ul,r}+2g_{u[i,j]}\big)} \,, \\
&& R_{ruij} = {\textstyle g_{u[i,j],r}+\Theta\big(g_{u[i}g_{j]u,r}-2g_{u[i,j]}\big)} \,, \\
&& R_{ruui} = {\textstyle g_{u[u,i],r}+\frac{1}{4}g^{rl}g_{ul,r}g_{ui,r}-\frac{1}{2}g^{kl}g_{uk,r}E_{li}} \nonumber \\
&& \hspace{15.0mm} {\textstyle +\Theta\big(g_{ui,u}-\frac{1}{2}g_{uu,i}-\frac{1}{2}g_{ui}g_{uu,r}\big)} \,, \\
&& R_{kilj} = {\textstyle (\frac{1}{2}\,^{S}\!R-\Theta^2g^{rr})(g_{kl}g_{ij}-g_{kj}g_{il})} \nonumber \\
&& \hspace{15.0mm} {\textstyle -\Theta\big(g_{kl}e_{ij}+g_{ij}e_{kl}-g_{kj}e_{il}-g_{il}e_{kj}\big)} \,, \\
&& R_{uikj} = {\textstyle g_{i[k,u||j]}+g_{u[j,k]||i}+e_{i[k}g_{j]u,r}} \nonumber \\
&& \hspace{15.0mm} {\textstyle +\Theta\big(g^{rr}g_{i[k}g_{j]u,r}+g_{uu,[j}g_{k]i}-2g^{rl}E_{l[j}g_{k]i}\big)} \,, \\
&& R_{uiuj} = {\textstyle -\frac{1}{2}g_{uu||ij}+g_{u(i,u||j)}-\frac{1}{2}g_{ij,uu}+\frac{1}{4}g^{rr}g_{ui,r}g_{uj,r}} \nonumber \\
&& \hspace{15.0mm} {\textstyle -\frac{1}{2}g_{uu,r}e_{ij}+\frac{1}{2}g_{uu,(i}g_{j)u,r}-g^{rl}E_{l(i}g_{j)u,r}+g^{kl}E_{ki}E_{lj}} \nonumber \\
&& \hspace{15.0mm} {\textstyle -\frac{1}{2}\Theta g_{ij}\big[g^{rr}g_{uu,r}+g_{uu,u}-g^{rl}(2g_{ul,u}-g_{uu,l})\big]} \,,
\end{eqnarray}
the components of the Ricci tensor are
\begin{eqnarray}
&& R_{rr} = {\textstyle -2\big(\Theta_{,r}+\Theta^2\big)} \,, \label{Ricci rr}\\
&& R_{ri} = {\textstyle -\frac{1}{2}g_{ui,rr}+g_{ui}\Theta_{,r}-\Theta_{,i}+2\Theta^2g_{ui}} \,, \label{Ricci rp} \\
&& R_{ru} = {\textstyle -\frac{1}{2}g_{uu,rr}+\frac{1}{2}g^{ri}g_{ui,rr}+\frac{1}{2}g^{ij}\big(g_{ui,r||j}+g_{ui,r}g_{uj,r}\big)} \nonumber \\
&& \hspace{12.0mm} {\textstyle -2\Theta_{,u}-\frac{1}{2}\Theta\big(g^{ij}g_{ij,u}+2g_{uu,r}\big)} \,, \label{Ricci ru} \\
&& R_{ij} = {\textstyle \frac{1}{2}\,^{S}\!R\, g_{ij}-g_{u(i,r||j)}-\frac{1}{2}g_{ui,r}g_{uj,r}-g_{ij}\big(g^{rr}\Theta_{,r}-2\Theta_{,u}+2g^{rl}\Theta_{,l}\big)+2g_{u(i}\Theta_{,j)}} \nonumber \\
&& \hspace{12.0mm} {\textstyle +\Theta^2\big(2g_{ij}g^{rl}g_{ul}-2g_{ij}g^{rr}-2g_{ui}g_{uj}\big)} \nonumber \\
&& \hspace{12.0mm} {\textstyle +\Theta\big[2g_{u(i||j)}+2g_{u(i}g_{j)u,r}-2e_{ij}+g_{ij}\big(g_{uu,r}-2g^{rl}g_{ul,r}-g^{kl}e_{kl}\big)\big]} \,, \label{Ricci pq} \\
&& R_{ui} = {\textstyle -\frac{1}{2}g^{rr}g_{ui,rr}-\frac{1}{2}g_{uu,ri}+\frac{1}{2}g_{ui,ru}+g^{rl}g_{u[l,i],r}-\frac{1}{2}g^{rl}(g_{ui,r||l}+g_{ul,r}g_{ui,r})} \nonumber \\
&& \hspace{12.0mm} {\textstyle +g^{kl}\big(\frac{1}{2}g_{uk,r}g_{ul||i}+g_{k[i,u||l]}+g_{u[k,i]||l}-\frac{1}{2}e_{kl}g_{ui,r}\big)}\nonumber \\
&& \hspace{12.0mm} {\textstyle +g_{ui}\Theta_{,u}+\Theta\big[g_{ui}g_{uu,r}-g_{ui,u}} \nonumber \\
&& \hspace{20.0mm} {\textstyle -g^{rl}g_{ul,r}g_{ui}-2g^{rl}(g_{u[l,i]}-\frac{1}{2}g_{ul}g_{ui,r})+g^{rl}g_{li,u}\big]}\,, \label{Ricci up} \\
&& R_{uu} = {\textstyle -\frac{1}{2}g^{rr}g_{uu,rr}-g^{rl}g_{uu,rl}-\frac{1}{2}g^{kl}e_{kl}g_{uu,r}+g^{rl}g_{ul,ru}-\frac{1}{2}g^{kl}g_{kl,uu}} \nonumber \\
&& \hspace{12.0mm} {\textstyle +g^{kl}(g_{uk,u||l}-\frac{1}{2}g_{uu||kl})+\frac{1}{2}(g^{rr}g^{kl}-g^{rk}g^{rl})g_{uk,r}g_{ul,r}} \nonumber \\
&& \hspace{12.0mm} {\textstyle +2g^{kl}g^{ri}g_{uk,r}g_{u[l,i]}+\frac{1}{2}g^{kl}g_{uk,r}g_{uu,l}+g^{kl}g^{ij}E_{ik}E_{jl}} \nonumber \\
&& \hspace{12.0mm} {\textstyle +\Theta\big(g_{uu}g_{uu,r}-g_{uu,u}\big)} \,, \label{Ricci uu}
\end{eqnarray}
and the Ricci scalar is
\begin{eqnarray}
&& R = {\textstyle \,^{S}\!R+g_{uu,rr}-2g^{rl}g_{ul,rr}-2g^{ij}g_{ui,r||j}-\frac{3}{2}g^{ij}g_{ui,r}g_{uj,r}} \nonumber \\
&& \hspace{8.0mm} {\textstyle +2\Theta_{,r}\big(2g_{uu}-g^{rl}g_{ul}\big)+8\Theta_{,u}-4g^{rl}\Theta_{,l}+6\Theta^2 g_{uu}} \nonumber \\
&& \hspace{8.0mm} {\textstyle +\Theta\big(4g_{uu,r}-2g^{rl}g_{ul,r}+3g^{ij}g_{ij,u}-2g^{ij}g_{ui||j}\big)} \,.
\end{eqnarray}

The Weyl tensor components are
\begin{eqnarray}
&& C_{rirj} = 0 \,, \label{Weyl rprq} \\
&& C_{riru} = {\textstyle \frac{1}{4}(-G_{ui,r}+2\Theta_{,i})} \,, \\
&& C_{rikj} = {\textstyle -\frac{1}{2}g_{i[k}G_{j]u,r}+g_{i[k}\Theta_{,j]}} \,,\\
&& C_{ruru} = {\textstyle -\frac{1}{3}\big[\frac{1}{2}\,^{S}\!R +\frac{1}{2}G_{uu,r} +\frac{1}{2}g^{ij}G_{ui||j}+\frac{1}{2}g^{ri}(G_{ui,r}-2\Theta_{,i})-2\Theta_{,u}\big]} \,, \\
&& C_{riuj} = {\textstyle \frac{1}{2}\big[\frac{1}{6}\,g_{ij}\,^{S}\!R+G_{u[i||j]} +\frac{1}{6}g_{ij}\big(G_{uu,r}+g^{rl}G_{ul,r}+g^{kl}G_{uk||l}\big)-\frac{1}{2}g_{ui}G_{uj,r}} \nonumber \\
&& \hspace{18.0mm} {\textstyle+g_{ui}\Theta_{,j}-\frac{2}{3}g_{ij}\Theta_{,u}-\frac{1}{3}g_{ij}g^{rl}\Theta_{,l}\big]} \,, \\
&& C_{ruij} = {\textstyle G_{u[i||j]}-\frac{1}{2}g_{u[i}G_{j]u,r} +g_{u[i}\Theta_{,j]}} \,, \\
&& C_{kilj} = {\textstyle
\frac{1}{6}\,(g_{kl}g_{ij}-g_{kj}g_{il})\big[\,^{S}\!R+G_{uu,r}-2g^{mn}G_{um||n}
-2g^{rn}(G_{un,r}-2\Theta_{,n})}\\
&& \hspace{42.0mm} {\textstyle -\frac{3}{2}g^{mn}G_{um}G_{un}-4\Theta_{,u} \big]} \nonumber\\
&& \hspace{12.0mm} {\textstyle
 +\frac{1}{4}g_{kl}\big(2G_{u(i||j)}+G_{ui}G_{uj} \big)
 +\frac{1}{4}g_{ij}\big(2G_{u(k||l)}+G_{uk}G_{ul} \big)} \nonumber \\
&& \hspace{12.0mm} {\textstyle
 -\frac{1}{4}g_{kj}\big(2G_{u(i||l)}+G_{ui}G_{ul} \big)
 -\frac{1}{4}g_{il}\,\big(2G_{u(k||j)}+G_{uk}G_{uj} \big)}  \,, \\
&& C_{ruui} = {\textstyle \frac{1}{2}G_{u[u,i]}+\frac{1}{4}g^{kl}G_{uk}(g_{ui||l}-g_{il,u})-\frac{1}{4}g^{kl}e_{kl}G_{ui}-\frac{1}{4}g^{rl}g_{ul}G_{ui,r}} \nonumber \\
&& \hspace{12.5mm} {\textstyle +\frac{1}{2}g^{kl}\big(g_{k[i,u||l]}+g_{u[k,i]||l}\big)+\frac{1}{4}g^{rl}\big(3G_{u[l||i]}-G_{u(i||l)}\big)+\frac{1}{2}g^{rl}g_{ul}\Theta_{,i}} \nonumber \\
&& \hspace{12.5mm} {\textstyle -\frac{1}{6}g_{ui}\big[\,^{S}\!R-\frac{1}{2}G_{uu,r}-\frac{1}{2}\big(g^{rl}G_{ul,r}+g^{kl}G_{uk||l}\big)+2\Theta_{,u}+g^{rl}\Theta_{,l}\big]} \,,
\end{eqnarray}
\begin{eqnarray}
&& C_{uikj} = {\textstyle g_{i[k,u||j]}+g_{u[j,k]||i}+e_{i[k}G_{j]u}-\frac{1}{2}\big(G_{ui||[k}g_{j]u}+g_{u[j}G_{k]u||i}+G_{ui}G_{u[k}g_{j]u}\big)} \nonumber \\
&& \hspace{12.5mm} {\textstyle +\frac{1}{2}\big[-g^{rr}g_{i[k}G_{j]u,r}-G_{uu,[j}g_{k]i}+2g^{rr}\Theta_{,[j}g_{k]i}+g_{i[k}G_{j]u,u}} \nonumber \\
&& \hspace{20mm} {\textstyle +g^{rl}\big(G_{ul||[j}g_{k]i}-2g_{i[k}G_{j]u||l}-G_{ul}g_{i[k}G_{j]u}\big)} \nonumber \\
&& \hspace{20mm} {\textstyle +g^{ln}\big(G_{ul}g_{un||[j}g_{k]i}+g_{ik}g_{l[j,u||n]}-g_{ij}g_{l[k,u||n]}} \nonumber \\
&& \hspace{30mm} {\textstyle +g_{ik}g_{u[l,j]||n}-g_{ij}g_{u[l,k]||n}-e_{ln}g_{i[k}G_{j]u}\big)\big]} \nonumber \\
&& \hspace{12.5mm} {\textstyle +\frac{1}{3}\,g_{i[k}g_{j]u}\big(\frac{1}{2}\,^{S}\!R+4\Theta_{,u}-4g^{rl}\Theta_{,l}} \nonumber \\
&& \hspace{32.5mm} {\textstyle -G_{uu,r}+2g^{rl}G_{ul,r}+\frac{3}{2}g^{ln}G_{ul}G_{un}+2g^{ln}G_{ul||n}\big)} \,, \\
&& C_{uiuj} = {\textstyle -\frac{1}{2}g_{uu||ij}-\frac{1}{2}g_{ij,uu}+g_{u(i,u||j)}-\frac{1}{2}G_{uu}e_{ij}+\frac{1}{2}g_{uu,(i}G_{j)u}+g^{mn}E_{mi}E_{nj}} \nonumber \\
&& \hspace{12.5mm} {\textstyle -\frac{1}{2}\,g_{ij}g^{kl}\big(-\frac{1}{2}g_{uu||kl}-\frac{1}{2}g_{kl,uu}+g_{uk,u||l}} \nonumber \\
&& \hspace{31.5mm} {\textstyle -\frac{1}{2}G_{uu}e_{kl}+\frac{1}{2}g_{uu,k}G_{ul}+g^{mn}E_{mk}E_{nl}\big)} \nonumber \\
&& \hspace{12.5mm} {\textstyle +\frac{1}{6}\,(g_{uu}g_{ij}-g_{ui}g_{uj})\big(\,^{S}\!R+G_{uu,r}-2g^{rl}G_{ul,r} -\frac{3}{2}g^{kl}G_{uk}G_{ul}-2g^{kl}G_{uk||l}\big)} \nonumber \\
&& \hspace{12.5mm} {\textstyle -\frac{1}{4}\,g_{uu}g_{ij}\big(\,^{S}\!R+G_{uu,r}-g^{kl}G_{uk}G_{ul}\big)} \nonumber \\
&& \hspace{12.5mm} {\textstyle +\frac{1}{4}g^{rl}g_{ul}G_{ui}G_{uj} +\frac{1}{2}\,g_{uu}G_{u(i||j)}-g^{rl}E_{l(i}G_{j)u}} \nonumber \\
&& \hspace{12.5mm} {\textstyle +\frac{1}{2}\,g_{ij}g^{rl}\big[\frac{1}{2}g_{ul}G_{uu,r}+G_{uu,l}-G_{ul,u}} \nonumber \\
&& \hspace{30.0mm} {\textstyle -\frac{1}{2}G_{um}\big(g^{mn}G_{un}g_{ul}-g^{rm}G_{ul}-4g^{mn}g_{u[l,n]}\big)\big]} \nonumber \\
&& \hspace{12.5mm} {\textstyle +\frac{1}{2}\big(-g^{rr}g_{u(j}G_{i)u,r}-G_{uu,(i}g_{j)u}+g_{u(j}G_{i)u,u}} \nonumber \\
&& \hspace{20.0mm} {\textstyle +g^{rl}G_{ul||(i}g_{j)u}-2g^{rl}g_{u(j}G_{i)u||l}-g^{rl}G_{ul}g_{u(j}G_{i)u}\big)} \nonumber \\
&& \hspace{12.5mm} {\textstyle +\frac{1}{2}\,g^{kl}\big[G_{uk}g_{ul||(i}g_{j)u}-e_{kl}g_{u(j}G_{i)u}+g_{u(j}g_{i)k,u||l}-g_{kl,u||(i}g_{j)u}} \nonumber \\
&& \hspace{26.0mm} {\textstyle +\frac{1}{2}\big(g_{uj}g_{uk||il}+g_{ui}g_{uk||jl}\big)-g_{u(j}g_{i)u||kl}\big]} \nonumber \\
&& \hspace{12.5mm} {\textstyle +\frac{1}{3}\Theta_{,u}\big(g_{uu}g_{ij}+2g_{ui}g_{uj}-3g^{rl}g_{ul}g_{ij}\big)} \nonumber \\
&& \hspace{12.5mm} {\textstyle +g^{rr}g_{u(i}\Theta_{,j)}+\frac{2}{3}g^{rl}\Theta_{,l}\big(g_{uu}g_{ij}-g_{ui}g_{uj}\big)} \,. \label{Weyl upuq}
\end{eqnarray}

\end{document}